\documentclass[12pt]{article}
\usepackage{amsmath}
\usepackage{amssymb}
\usepackage{graphicx}
\usepackage{epsfig}
\usepackage{color}
\usepackage{fullpage}
\usepackage{url}

\newcommand\pubdate{}

\def\STANKI{Department  of Physics,\\
 Massachusetts Institute of Technology , Cambridge, MA 02139}
 \def\YALE{Department  of Physics,  Yale University \\
P.O. Box 208120, New Haven, CT 06520}

\def\Title#1{\begin{center} {\Large #1 } \end{center}}
\def\Author#1{\begin{center}{ \sc #1} \end{center}}
\def\Address#1{\begin{center}{ \it #1} \end{center}}

\newcommand\pubblock{\rightline{\begin{tabular}{l}\\
         \pubdate \end{tabular}}}

\newcommand{\bp}{\bar{\varphi}}

\newcommand{\mpl}{m_{\textrm{pl}}}

\newcommand{\tk}{\tilde{k}}
\newcommand{\nl}{\textrm{nl}}
\newcommand{\p}{\textrm{p}}
\newcommand{\z}{(\lambda/g)}

\date{ }
\begin{document}

\begin{titlepage}
\pubblock

\vfill \Title{Inflaton Fragmentation and Oscillon Formation \\in  Three Dimensions  }
\vfill \Author{Mustafa A. Amin\footnote{email: mamin@mit.edu}, Richard Easther\footnote{email: richard.easther@yale.edu} \& Hal Finkel\footnote{email: hal.finkel@yale.edu} } 
\Address{\STANKI} 
\Address{\YALE}

\begin{abstract}
Analytical arguments suggest that  a large class of scalar field potentials permit the existence of oscillons --   pseudo-stable, non-topological solitons -- in three spatial dimensions.  In this paper we  numerically explore oscillon solutions in three dimensions.  We confirm the existence of these field configurations as solutions to the Klein-Gorden equation in an expanding background,  and verify the predictions of Amin and Shirokoff \cite{Amin:2010jq} for the characteristics of individual oscillons for their model. Further, we demonstrate that significant numbers of oscillons can be generated via fragmentation of the inflaton condensate, consistent with the analysis of Amin \cite{Amin:2010xe}. These emergent oscillons can easily dominate the post-inflationary universe. Finally, both analytic and numerical results suggest that oscillons are  stable on timescales longer than the post-inflationary Hubble time. Consequently, the post-inflationary universe can contain an effective matter-dominated phase, during which it is dominated by localized concentrations of scalar field matter.
\end{abstract}

\vfill
\end{titlepage}
\def\thefootnote{\fnsymbol{footnote}}
\setcounter{footnote}{0}

\section{Introduction}

Oscillons are long-lived, oscillatory, spatially localized solutions that exist in a large class of real-valued, scalar field models with nonlinear interactions \cite{Amin:2010jq,Amin:2010xe,Bogolyubsky:1976yu,Gleiser:1993pt, Honda:2001xg, Adib:2002ff, Farhi:2005rz, Gleiser:2006te, Graham:2006xs, Hindmarsh:2006ur, Fodor:2006zs, Saffin:2006yk, Graham:2006vy, Arodz:2007jh, Hindmarsh:2007jb, Fodor:2008es, Gleiser:2008ty, Fodor:2008du, Fodor:2009xw, Gleiser:2009ys, Gleiser:2007te,  Hertzberg:2010yz}, and can arise from generic initial field configuations \cite{Gleiser:2003uu,Copeland:1995fq, Farhi:2007wj, Amin:2010xe, Dymnikova:2000dy, McDonald:2001iv, Broadhead:2005hn,Copeland:2002ku, Kasuya:2000wx}.    Consequently, it is natural to ask whether oscillons are found in the post-inflationary universe, which is typically dominated by real-valued field(s) with nonlinear interactions.   This question has two parts. Firstly, we need to confirm whether oscillonic solutions actually exist for potentials corresponding to inflationary scenarios. Secondly, given the existence of these solutions, we can ask if the post-inflationary dynamics naturally leads to the generation of oscillons.  In this paper, we numerically confirm previous analytic estimates of oscillon properties, showing that both questions can be answered in the affirmative.  

From the cosmological perspective, it has long been known that the post-inflationary universe can host a vast array of strongly nonlinear processes, thanks to parametric resonance and (p)reheating \cite{Traschen:1990sw,Kofman:1994rk,Shtanov:1994ce,Kofman:1997yn, Podolsky:2005bw,GarciaBellido:2002aj, Felder:2001kt, Felder:2000hj, Tkachev:1998dc, Khlebnikov:1996mc,Greene:1998pb,Dufaux:2006ee}.   Resonance fragments the inflaton condensate, rendering the universe strongly inhomogeneous on sub-horizon scales, and our work will make it clear oscillon formation is one of a long list of phenomena associated with resonance (also see \cite{Amin:2010xe, McDonald:2001iv, Broadhead:2005hn, Copeland:2002ku,Kasuya:2000wx}).  Resonance is usually linked to reheating and thermalization, but oscillons are effectively large, long-lived ``blobs'' of scalar field matter, and thus provide a mechanism by which resonance can be followed by an effective matter-dominated phase.   

Oscillons are not exactly stable, as they are not associated with  a strictly conserved charge in the underlying theory.  In practice, however, they decay on timescales much longer than  $m^{-1}$,  where $m$ is the mass in the scalar field Lagrangian \cite{Segur:1987mg,Fodor:2009kf,Gleiser:2009ys,Hertzberg:2010yz,Farhi:2007wj}.   However, for oscillons to have significant cosmological consequences their lifetime should also exceed the  Hubble time at formation ($H^{-1}$), an interval which is also typically much longer than $m^{-1}$.  Interestingly, while oscillonic solutions have long been known, these solutions typically  have relatively small amplitudes and suffer from a collapse instability against long-wavelength perturbations, with the result that their typical lifetime $\tau_{osc}$ falls in the range $m^{-1} \lesssim \tau_{osc} \lesssim H^{-1}$.   Recently, Amin and Shirokoff \cite{Amin:2010jq} described a new class of oscillon solutions, corresponding to ``flat-topped" configurations with large masses, and these objects are stable against long-wavelength perturbations.  Further, in \cite{Amin:2010xe}, Amin showed that these oscillon solutions can be found in potentials that also undergo post-inflationary parametric resonance, which drives the formation of oscillons. 

Given that the mass and lifetime of the flat-top oscillons  exceeds that of their small-amplitude, ``Gaussian" cousins  \cite{Amin:2010jq,Amin:2010xe},  they provide a recipe for constructing oscillon-dominated cosmological scenarios.   Our starting point is  the potential  %
\begin{equation} \label{eq:potential}
V(\varphi)=\frac{1}{2}m^2\varphi^2-\frac{\lambda}{4}\varphi^4+\frac{g^2}{6m^2}\varphi^6
  \end{equation}
with $\lambda>0$ and $\z^2\ll1$.   A necessary condition for  the existence of oscillons is that the potential ``opens out'' as $\varphi$ increases from zero, hence $\lambda>0$. The $\varphi^6$ term is required for the flat-top solutions, and the underlying analysis  makes use of an expansion in $\z$ \cite{Amin:2010jq,Amin:2010xe}. Although the linear stability of the oscillon solutions was investigated in \cite{Amin:2010jq}, it is important to investigate the robustness of oscillon solutions numerically, given that this system is intrinsically non-linear. In addition, although \cite{Amin:2010xe} provided semi-analytic results for the post inflationary production of oscillons in three dimensions, a detailed numerical study of the fully non-linear evolution was performed only in $1+1$ dimensions. Consequently, it is clear that we need to numerically  confirm both the existence and generation of oscillons in three spatial dimensions  before proceeding to an analysis of their consequences.  

We note that the potential considered here is not a viable model of inflation, since $V(\varphi) \propto \varphi^6$ at large $\varphi$, and a sixth order slow roll potential is clearly at odds with the data.  Rather, $V(\varphi)$ is taken to be a truncated Taylor series that describes the potential at small field values, or $\varphi\ll\mpl$, and is applicable only after inflation ends. The inflationary portion of the potential (typically $\varphi\gtrsim \mpl$) is not accurately captured by the truncated expression.   We are further assuming that the self couplings are more important than couplings to other fields during the era of interest. The hard lower limit on the thermalization temperature is set by the generation of a cosmological neutrino background and successful nucleosynthesis \cite{Alpher:1948,Wagoner:1966pv} while inflation is typically a (near) GUT scale phenomenon. Consequently, while these couplings must be non-zero they can still be very small.  However, it is easy to imagine scenarios in  which couplings to other fields drain the energy from the inflaton condensate before oscillons form. In this paper, however, we focus on the generic, truncated potential and will discuss oscillon formation in realistic inflationary scenarios including the impact of couplings between the inflaton and other fields  in a future paper.   We will further restrict out attention to scenarios where $\z^2 \lesssim 10^{-1}$, which corresponds to insisting that the $\varphi^6$ term in the Taylor expansion is large. In realistic scenarios where this condition is not satisfied, oscillon formation is still entirely possible, but the solutions may not be well described by the approximately single frequency solutions found when $\z^2 \ll 1$. Finally, this analysis is performed in an expanding universe but ignores ``local'' gravitational interactions associated with the oscillons, which can potentially lead to primordial blackhole formation \cite{Khlopov:2008qy} or gravitational wave production  \cite{Khlebnikov:1997di, Easther:2006gt,Dufaux:2007pt,Easther:2006vd, Easther:2007vj, GarciaBellido:2007af,Giblin:2010sp}, and these topics will also be pursued in future work.

  \begin{figure}[bt]  
     \centering
     \includegraphics[width=4in]{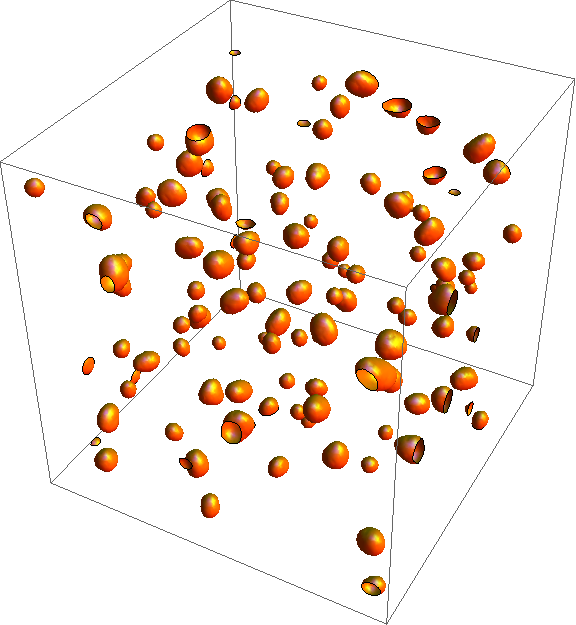} 
     \caption{Sample output from a 3D oscillon simulation. Contours delineate regions where the local energy  density exceeds the average value by a factor of 10, showing localized, approximately spherical, overdensities characteristic of oscillons. The comoving simulation volume contains a significant fraction of a single, post-inflationary Hubble volume.}
     \label{fig:sample}
  \end{figure}

Oscillon-like solutions  in both the post-inflationary universe and more general  nonlinear theories have been widely studied.  Oscillon formation with quasi-thermal initial conditions in a $1+1$ dimensional de-Sitter universe was investigated in \cite{Farhi:2007wj}, while  oscillon-oscillon and oscillon-domain wall interactions in $2+1$ dimensions are discussed in \cite{Hindmarsh:2007jb}. Q-balls \cite{Coleman:1985ki,Lee:1991ax,Enqvist:2002rj,Enqvist:2002si,Kusenko:2008zm,Kusenko:2009cv,Chiba:2009zu}  have many similarities with oscillons, but differ in that they arise in complex valued fields with conserved charges. Conversely, the oscillons discussed here exist in a real valued field with no exactly conserved charge, although we do  have an adiabatic invariant \cite{Kasuya:2002zs}. Likewise, the formation of ``axitons" in the axion field during the QCD phase transition was investigated in \cite{Kolb:1993hw} and pseudo-solitons (which are not necessarily oscillons) have been investigated in certain supersymmetric (SUSY) hybrid inflation models \cite{Copeland:2002ku, Kasuya:2000wx}. A very detailed numerical study of $Q$-ball formation in gravity mediated SUSY breaking models is given by \cite{Hiramatsu:2010dx}. An analytic and numerical investigation ($64^3$ lattice) of inflaton fragmentation into oscillons was carried out in  \cite{McDonald:2001iv, Broadhead:2005hn} for a class of supersymmetric models. For the model described by equation~\eqref{eq:potential}, Amin \cite{Amin:2010xe} provided semi-analytic predictions for the post-inflationary number density of oscillons in $3+1$ and $1+1$ dimensions while the full non-linear field evolution was investigated in $1+1$ dimensions.  In these $1+1$ dimensional simulations, oscillons comprised $\sim 80\%$ of the total energy density at late times. Finally, Gleiser and collaborators treated  oscillon production in the early universe from quasi-thermal initial conditions, reporting a $1.2-3\%$ fraction of energy density in oscillons  for a double-well model and an $SU(2)$ model \cite{Gleiser:2010qt}, in contrast to the fully oscillon dominated phase ($\sim 75\%$) we describe here.  

Numerical simulations of oscillons in $3+1$ dimensions are challenging: oscillons are small, localized objects of fixed {\em physical\/} size. Conversely, they are generated by parametric resonance at comoving scales that redshift relative to the oscillon radius as the universe expands, so any simulation must resolve a broad range of spatial scales, with good stability over a number of Hubble times.   Further, the properties of  oscillons depend nontrivially on the  number of spatial dimensions, so we must work with a full 3D code, without any symmetry assumptions. The simulations described here were performed using two different codes, a modified version of {\sc Defrost} \cite{Frolov:2008hy},  and {\sc PSpectRe}  \cite{Easther:2010qz}.\footnote{We added support for a sixth order single field potential to  {\sc Defrost}, and adapted it to run in an MPI-based cluster environment. {\sc PSpectRe} and the modified {\sc Defrost} are available for download at \url{http://easther.physics.yale.edu/downloads.html}. The visualizations of our simulations are available at  \url{http://www.mit.edu/~mamin/oscillons.html} and \url{http://easther.physics.yale.edu/fields.html}.}  {\sc Defrost} uses a finite differencing scheme, whereas  {\sc PSpectRe} is a pseudo-spectral code in which the fundamental degrees of freedom are the momentum modes of scalar field(s), and the two codes agree well.    A representative output from one of our simulations showing the oscillon distribution well after transients associated with the initial formation phase is shown in Figure~\ref{fig:sample}.

This paper is  organized as follows.  Oscillon solutions and their stability are discussed in Section \ref{sec:model}. In Section \ref{sec:icl} we review the amplification of zero point fluctuations in the inflaton field by parametric resonance, and estimate the resulting number density of oscillons.  Section \ref{sec:ns3} contains the main new results in this paper. There, we provide numerical results and compares them to the theoretical estimates. Finally, we conclude with a summary of our results and possible future directions. 

\section{Oscillons:   Analytic Expectations}
\label{sec:model}

We begin by summarizing the analytic expectations for oscillon solutions in three dimensions  as described in \cite{Amin:2010jq} before numerically verifying these results.  The inflaton action  is 
\begin{equation}
 {S}=\int dx^4\sqrt{-\mathcal{G}} \left[\frac{\mpl^2}{2}R-\frac{1}{2}\partial^\mu\varphi\partial_\mu\varphi-V(\varphi)\right],
\end{equation}
where $\mathcal{G}$ is the determinant of the metric, $R$ is the Ricci scalar, $\mpl$ is the reduced Planck mass and $\hbar=c=1$. The potential $V(\varphi)$ is given by equation~\eqref{eq:potential} near its minimum:  $m$ is the effective mass of the inflaton at the bottom of the potential,  while $\lambda$ and $g$ are dimensionless parameters.  At large values of $\varphi\gtrsim \mpl$, the potential must be relatively flat in order to support inflation. We do not make a direct connection to detailed inflationary phenomenology at this point.    

    \begin{figure}[tbp] 
     \centering
     \includegraphics[width=6in]{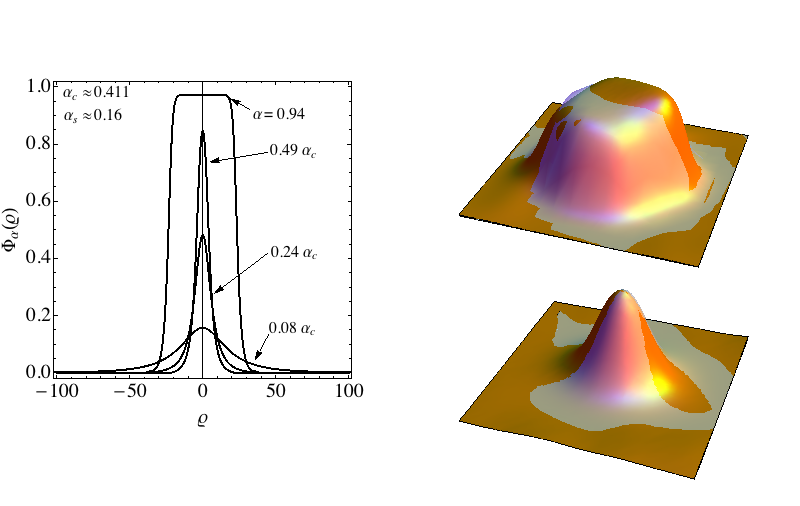} 
     \caption{The left panel shows the theoretically calculated radial field profiles of oscillons from Amin and Shirokoff \cite{Amin:2010jq}. The right panel shows the radial cross-section of the energy density of two oscillons from a fiducial run showing the flat topped, as well the Gaussian oscillon. The white surface represents our one parameter fit based on the theoretically calculated profiles.}
     \label{fig:flattops}
  \end{figure}

The  field $\varphi$ obeys the usual Klein-Gordon equation,
 \begin{equation}
 \begin{aligned}
\Box \varphi=V'(\varphi).
 \end{aligned}
 \end{equation} 
Assuming spherical symmetry and ignoring expansion (for now) gives 
\begin{equation}
 \partial_t^2{\varphi}-\partial_r^2\varphi-\frac{2}{r}\partial_r\varphi+m^2\varphi-\lambda\varphi^3+\frac{g^2}{m^2}\varphi^5=0 \, .
 \end{equation}
Spatially localized solutions generically require a potential  that is ``shallower'' than quadratic near its minimum. Given equation \eqref{eq:potential}, this implies $\lambda>0$ and we {\em assume\/}  that $(\lambda/g)^2\ll1$. This latter constraint is not required, but is the basis of the semi-analytic treatment of the flat-top solutions \cite{Amin:2010jq}.  For further convenience, we take $\lambda,g\ll1$ and $m/\mpl\ll1$. The fiducial values we have in mind are $m/\mpl\sim 5\times 10^{-6}$ and  $\lambda,g\sim 10^{-6}$, while typical field values satisfy $\varphi\lesssim m/\sqrt{\lambda}$.    

We are interested in field configurations that are spatially localized and periodic in time, and start with the following ansatz for a one parameter family of solutions indexed by $\alpha$ is 
\begin{equation}
\varphi(t,r)\approx \frac{m}{\sqrt{\lambda}}\left[\left(\frac{\lambda}{g}\right)\Phi_{\alpha}(\varrho)\cos\omega t  + \mathcal{O}[(\lambda/g)^3]\right],
\end{equation}
where we define $\omega^2=m^2-(\lambda/g)^2\alpha^2$ and $\varrho=(\lambda/gm)r$ \cite{Amin:2010jq}. The expansion in $\z$ is similar to the often used $\epsilon$ expansion (e.g. \cite{Fodor:2008du}), but can access the entire allowed range of harmonic, spatially localized solutions. Requiring  that $\varphi(t,r)$ is both periodic and bounded can be shown to lead to the requirement that the  spatial envelope satisfies 
 \begin{equation}
 \label{eq:3dprofile}
\partial_\varrho^2\Phi_\alpha+\frac{2}{\varrho}\partial_\varrho\Phi_\alpha-\alpha^2\Phi_\alpha+\frac{3}{4}\Phi_\alpha^3-\frac{5}{8}\Phi_\alpha^5=0 \, ,
\end{equation}
which is the equation for the oscillon profile  \cite{Amin:2010jq}.  

Assuming smoothness at the origin and spatial localization provides boundary conditions at $r=0$ and $r\rightarrow \infty$. The profile depends smoothly on $\alpha$, which must lie between $0< \alpha <\alpha_c$ for single-frequency, localized solutions where $\alpha_c=\sqrt{27/160}$. As shown in Figure \ref{fig:flattops}, $\alpha\ll\alpha_c$ gives small-amplitude Gaussian profiles, while  we find   wide, ``flat-topped" profiles as $\alpha\rightarrow\alpha_c$ \cite{Amin:2010jq}. At $\alpha_c$, the solution becomes homogenous in space, with amplitude $\Phi_{\alpha_c}=\sqrt{9/10}$.  

The stability of oscillons is determined by the slope of the energy of the oscillon vs. $\alpha^2$ curve \cite{Amin:2010jq} (also see \cite{Fodor:2009kf,Lee:1991ax}). The presence of the $\varphi^6$ in the potential allows this slope to change sign at 
$\alpha_s\approx 0.15$. In the $\alpha\ll\alpha_s$ limit (small-amplitude ``Gaussian" profile), $3+1$ dimensional oscillons suffer from a collapse instability when the wavelength of the perturbations is comparable to the width of the oscillons \cite{Amin:2010jq}. This instability disappears for $\alpha_s<\alpha<\alpha_c$. Consequently, we do not expect to see persistent solutions with $\alpha \lesssim   \alpha_s$.  Conversely, in the $\alpha\rightarrow\alpha_c$ limit, the oscillons are extremely wide, and  efficiently transfer energy to $(k\sim m)$ perturbations \cite{Amin:2010jq,Hertzberg:2010yz}. Wide oscillons are locally analogous to a homogeneous, oscillating field and can pump high frequency resonant modes (as described in the next section).  As a result, we expect an upper limit on the width of of the oscillons produced in our simulations.     
  
 Extracting the profile $\Phi_\alpha$ from numerical simulations is a nontrivial task, since the oscillons can be somewhat asymmetric and the field profile is modulated by the time dependent $\cos(\omega t)$ term. Conversely, the energy-density profiles, $\rho_\alpha$ does not vary significantly  with time, and is related to  $\Phi_\alpha$ by
\begin{equation}
\rho_{\alpha}(r)=m^4\frac{\lambda}{g^2}\left[\frac{1}{2}\Phi_{\alpha}^2+\left(\frac{\lambda}{g}\right)^2\left\{\frac{\left(\partial_{\rho}\Phi_{\alpha}\right)^2}{2}-\frac{1}{4}\Phi_{\alpha}^4+\frac{1}{6}\Phi_{\alpha}^6\right\}\right].
\end{equation}
The two critical values of $\rho_\alpha$, corresponding to $\alpha_c$ and $\alpha_s$, are 
\begin{equation}
\begin{aligned}
&\rho_c=m^4\left(9\lambda/20g^2\right)\left[1+\mathcal{O}[(\lambda/g)^2\right],\\
&\rho_s\approx   0.23 \frac{m^4\lambda}{g^2}.
\end{aligned}
\end{equation}
We expect to see long lived oscillons in the regime $\rho_{\alpha_s}<\rho_\alpha<\rho_{\alpha_c}$.

Given that we are assuming a radial profile, this analysis gives a match between the width (or, more conveniently, the cube root of the volume) and core density of the oscillons seen in our simulations, and this will be the relationship we seek to confirm numerically.   For convenience,  we take individual oscillons to occupy the region where the energy density is greater that $1/e$ of its value at the peak. Unlike most cases discussed in the literature, the width is a non-monotonic function of the core amplitude, with the increasing width at large core amplitudes being driven by the $\varphi^6$ term \cite{Amin:2010jq}. The relationship between the height and width of the oscillon energy densities is plotted in Figure \ref{fig:histhw}(a). 
Furthermore,   beyond $\rho_c$ the width is a two-valued function of the amplitude, but still approaches the homogeneous solution via the flat-top profiles. 
  
  When the expanding background is included, the solutions change character at $r^{*}\sim\alpha[(g/\lambda) H]^{-1}$, becoming oscillatory in space \cite{Amin:2010jq,Farhi:2007wj}. Thus, there is an upper limit on the oscillon width in an expanding universe, even though equation \eqref{eq:3dprofile} has solutions of arbitrary width. Physically,  oscillons with width $\gtrsim r^*$ are stretched  by the expansion while the flat space treatment  is reliable when the oscillon width obeys $r_e\lesssim r^{*}$. 
 Finally, the expanding background  slowly extracts energy from the oscillons in the form of outgoing radiation \cite{Farhi:2007wj, Graham:2006xs, Amin:2010jq}, in addition to the tiny, radiative loss not captured by a perturbative analysis \cite{Segur:1987mg}.
  
  \section{Resonance  and Oscillon Generation}
    \label{sec:icl}
 
 The formal existence of oscillon solutions is not sufficient to ensure that oscillons are actually generated in the very early universe.  Immediately following  inflation, the universe is homogeneous, with the inflaton oscillating about the minimum of $V(\varphi)$.    However, the same higher order terms in the potential that support the oscillon configurations also drive parametric resonance (see for example \cite{Traschen:1990sw,Kofman:1994rk,Shtanov:1994ce,Kofman:1997yn, Podolsky:2005bw,GarciaBellido:2002aj, Felder:2001kt, Felder:2000hj, Tkachev:1998dc, Khlebnikov:1996mc,Greene:1998pb,Dufaux:2006ee,Amin:2010xe}). Consequently, field perturbations at wavelengths within the resonance bands grow quasi-exponentially. For the potential here, parametric resonance in an expanding background and its consequences for oscillon formation were discussed by Amin in  \cite{Amin:2010xe}, along with an  approximate prediction for the resulting number density. We review the relevant results below before testing them numerically. 
  
 \subsection{Background and Resonance and  in the Linear Regime}
 
 While resonance is often studied in multifield models, it also occurs in single field models with non-quadratic potentials. The resonance bands for the potential  in equation \eqref{eq:potential} are described in \cite{Amin:2010xe}.   The homogeneous field obeys 
\begin{equation}
 \begin{aligned}
& \partial_{t}^2\bp+3H \partial_{t}\bp+V'(\bp)=0,\\
& H^2=\frac{1}{3\mpl^2}\left[ \frac{1}{2}\partial_{t}^2\bp+V(\bp)\right],
 \end{aligned}
 \end{equation}
 where $H$ is related to the scale factor $a$ in the usual way, $H=\dot{a}(t)/a(t)$.   At the bottom of the potential, we have
 \begin{equation}
 \begin{aligned}
&\bp(t)\approx \frac{\bp_i}{\sqrt{a^3(t)}}\cos(\omega t) \, ,\\
& H\approx\frac{H_i}{\sqrt{a^3(t)}} \, ,
 \end{aligned}
 \label{eq:homsol}
 \end{equation}
and $\omega^2\approx(m^2-\frac{3\lambda}{4}\bp^2+\hdots)$ when $\bp$ is small enough that the potential is dominated by the quadratic term. We define $a(t_i)=a_i=1, \bp(t_i)=\bp_i$, $H(t_i)=H_i$ and choose $\bp_i$ (and hence $t_i$)  to be the value of the homogenous field at which modes first enter the resonance band responsible for oscillon formation (see \cite{Amin:2010xe} for details),
 \begin{equation}
 \begin{aligned}
\label{eq:initialfield}
&\bp_i=\sqrt{\frac{3\lambda}{5g^2}} m \, . 
\end{aligned}
\end{equation}
During coherent oscillations, the universe is briefly potential-dominated during each cycle, with $\ddot{a}$ instantaneously positive.  Consequently, we chose the initial phase by setting  $\ddot{a}(t_i)=0$, or
\begin{equation}
\begin{aligned}
&H_i\approx\sqrt{\frac{3}{20}\frac{\lambda}{g^2}}\left(\frac{m}{\mpl}\right)m.
\end{aligned}
\end{equation}
 
 \subsection{Initial Conditions}
 
Given our choice of $t_i$ (via equation \eqref{eq:initialfield}) we begin our simulations well after the end of inflation, and we must provide initial conditions for  the field fluctuations, $\delta\varphi(t_i)$. We express these via a Fourier amplitude $\delta\varphi_k$, where $k$ is the {\em physical} wavenumber at $t_i$. Perturbations that leave the horizon during inflation are substantially modified relative to their original Bunch-Davies form, and evolve further following horizon re-entry after inflation \cite{Jedamzik:2010dq,Easther:2010mr}.  However, oscillon formation is dominated by modes that never left the horizon, and we set the initial perturbation spectrum consistent with the zero point fluctuations of the field. Treating these fluctuations as a classical, Gaussian, random field, a typical  fluctuation at a lengthscale $k^{-1}$ is given by (ignoring interaction terms):
\begin{equation}
\label{eq:icfluc}
k^{3/2}\langle|\delta\varphi_k(t_i)|^2\rangle^{1/2}\sim \frac{k^{3/2}}{\sqrt{2\omega_k}}
\end{equation}
where $\omega_k^2=k^2+m^2+\mathcal{O}[H_i^2]$ (see  \cite{QFTBook:2007}, for example), and $H$ is the Hubble parameter. Since we assume that $H_i\ll m$,  we will ignore the $H_i$ piece. We treat the fluctuations classically at all times, even though their occupation number is initially zero. However, parametric resonance amounts to particle production, and occupation numbers grow rapidly once resonance begins, justifying our use of classical field  theory for these simulations \cite{Khlebnikov:1996mc}.

Our potential also has narrow resonance bands for $\bp>\bp_i$ ($t<t_i$) that can change the initial conditions described above when $k\gtrsim m$. We evolved the linearized perturbations numerically for $t<t_i$  for some ``sample'' cases. Some modes with $k\gtrsim m$  are indeed amplified but this does not appear to have a significant effect on oscillon production or evolution. We also checked that our numerical simulations and qualitative results are robust against large changes ($\sim 10^2$) to the initial fluctuation amplitude.  From now on we  use \eqref{eq:icfluc} to specify initial conditions at $t_i$ and concentrate of the evolution of fluctuations after $t_i$.
 
\subsection{Resonance}
In  Fourier space, the linearized field fluctuations obey
 \begin{equation}
 \begin{aligned}
  \label{eq:LinearFourier}
 \partial_{t}^2\delta\varphi_{k}+3H \partial_{t}\delta\varphi_{k}+\left(\frac{k^2}{a^2(t)}+V''(\bp)\right)\delta\varphi_k
 =F_k\left[\delta\mathcal{G}_{\mu\nu}\right],
 \end{aligned}
 \end{equation}
where the right hand side is due to fluctuations of the metric, and the solution depends strongly on the relationship between $k, H$ and $V''(\bp)$. We assume $V''(\bp)-m^2\gg F\left[\delta\mathcal{G}_{\mu\nu}\right]$, and thus  ignore  the gravitational effects. We have checked that this assumption is valid over the length and time scales of interest for oscillon formation and for the parameter ranges considered here. The effects of field and gravitational instability in a related system in the post recombination universe was also discussed in \cite{Johnson:2008se}.

  \begin{figure}[tbp] 
     \centering
     \includegraphics[width=5.5in]{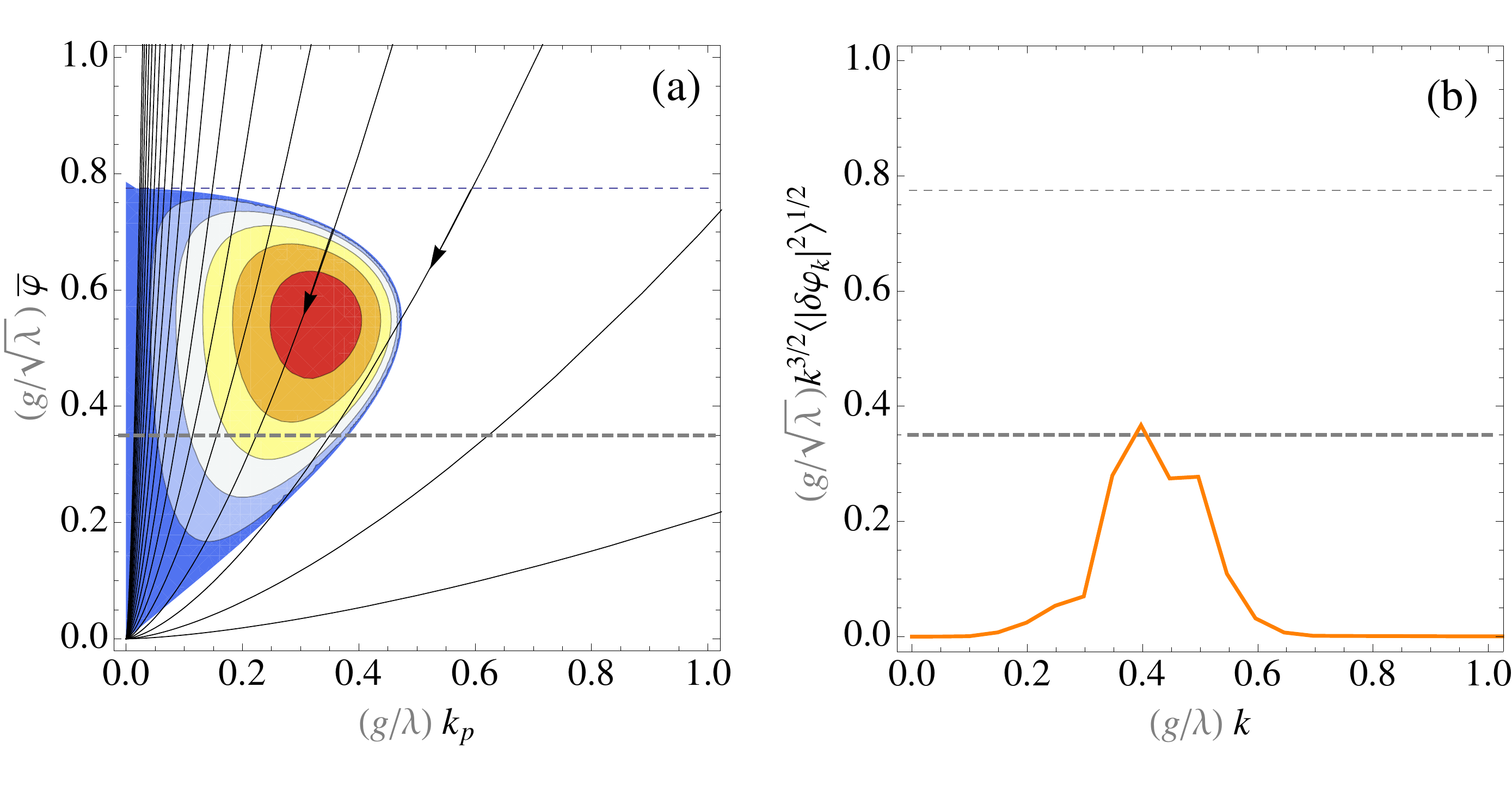} 
     \caption{In Figure (a), the resonance band for the potential $V(\bp)=m^2\bp^2/2-\lambda\bp^4/4+g^2\bp^6/6m^2$ is shown. The shaded region has a positive Floquet exponent. The thin, black lines show the ``path" following by modes as the universe expands.  Our simulations begin at the thin, dashed line. The thick, dashed indicates the homogenous field value where the evolution becomes non-linear, with fluctuations becoming comparable to the homogeneous field.  In Figure (b), the fluctuation power spectrum is shown at the point where linear analysis starts breaking down. The horizontal axis in Figure (a) is physical momentum, whereas in Figure (b) it is the co-moving momentum normalized at thin, dashed line. The momenta and field values are expressed in units of the mass, $m$. Figure (a) is adapted from \cite{Amin:2010xe}}
     \label{fig:floquet}
  \end{figure}

The instability bands and Floquet indices for the oscillon potential are derived in  \cite{Amin:2010xe}, and plotted in Figure \ref{fig:floquet}(a). The colors represent the real part of the Floquet exponent as a function of the field amplitude $\bp$ and physical wavenumber $k_{\textrm{p}}$, in flat space. In an expanding universe,  $\bp\approx\bp_ia^{-3/2}$ and $k_{\textrm{p}}=k a^{-1}$. As a result, a fluctuation with a given co-moving momentum samples different regions of this instability chart. Given that the expansion rate is slow compared to $m$, the amplification of a given comoving mode can be approximated by integrating the Floquet exponent along the thin black curves shown in Figure \ref{fig:floquet}(a). The band of wavenumbers that leads to oscillon formation is 
\begin{equation}
\frac{H_i}{\sqrt{a}}\ll k\lesssim 0.65\,\left(\frac{\lambda}{g}\right) m.
\end{equation}
The amplification undergone by these modes can be approximated by 
\begin{equation}
\begin{aligned}
\label{eq:fullsol}
\delta\varphi_k(a)\sim \frac{1}{\sqrt{2\omega_k}}\frac{1}{a^{3/2}}\exp\left[ \beta f(\tilde{k},a)\right],
\end{aligned}
\end{equation}
where
\begin{equation}
\begin{aligned}
&f(\tilde{k},a)=\sqrt{\frac{5}{3}}\int_{\frac{a^3-1}{a}\ge\frac{10}{9}\tilde{k}^2} d\ln \bar{a}\left[\tilde{k}\sqrt{\frac{9}{10\bar{a}^2}\left(1-\frac{1}{\bar{a}^3}\right)-\frac{\tilde{k}^2}{\bar{a}}}\right],\\
&\beta =\sqrt{\lambda}\frac{\lambda}{g}\left(\frac{\mpl}{m}\right),\\
&\tk = \frac{g}{\lambda m} k.
\end{aligned}
\end{equation}
Significant amplification requires $\beta\gg1,$  which provides an important constraint on the parameters $\lambda,g$ and $m$ \cite{Amin:2010xe}. The parameter $\beta\sim \mu/H$ where $\mu$ is the Floquet exponent characterizing the growth rate of fluctuations in flat space and $H$ is the Hubble parameter. Hence the condition $\beta\gg1$ follows from noting that to have rapid amplification of fluctuations, their growth rate $\mu$ must be larger that the Hubble rate $H$. Typical  resonant modes  have  $k\lesssim m\z$, and $\mu\sim 0.1m\z^2$. We have assumed $\z^2\sim 0.1$, while  $H\sim m\z^2/\beta\sim 10^{-3}m$.  For these values $\beta\sim 10^2$. The numerically-evaluated spectrum of fluctuations (linear regime) is shown in Figure \ref{fig:floquet}(b). The spectrum is in excellent agreement with the Floquet analysis,  justifying the approximations made in deriving the above expressions \cite{Amin:2010xe}.
 
We have concentrated on the first instability band and largely ignored the higher order bands (for example at $k_\p\sim\sqrt{3}m$).   The higher order bands are very narrow  ($\Delta k_\p \lesssim(\lambda/g)^2$) so modes redshift rapidly through these bands, with their maximum Floquet exponent scaling as $\mu\propto (\lambda/g)^2$.   Moreover, oscillons are significantly larger than $m^{-1}$ so we do not expect these high-$k_\p$ modes to significantly affect oscillon formation. Nevertheless, if modes pass through a large number of such bands or spend a long time in one of these narrow bands (that is, $H$ is very small) large amplification is possible and could lead to oscillon formation.  We checked numerically that these high-$k$ modes have little impact on oscillon formation for this scenario.\footnote{We show spectra in Figure \ref{fig:spectrum} --  the small bump  at $k_{\textrm{p}}\sim m^{-1}$ (for $a\lesssim2$) is produced by  the  the $k_\p\sim\sqrt{3}m$ band. We discuss  instability further at the end of Section \ref{sec:ns3}.}

 
   \subsection{Number density of oscillons}
 The linear analysis  breaks down when $\delta\varphi\sim k^{3/2}\langle|\delta\varphi_k|^2\rangle^{1/2}$ is comparable to $\bp$ as we no longer have a homogeneous pump field to parametrically amplify the fluctuations. In addition, as $\delta\varphi$ approaches $V''(\varphi)/V'''(\varphi)$, we have to take into account the interaction between different $k$ modes. The modes start interacting strongly with each other, ultimately forming oscillons. This process is difficult to follow analytically, but we can estimate the initial number density of oscillons, and its dependence on the free parameters in the potential \eqref{eq:potential} based on the linear analysis of growth of fluctuations \cite{Amin:2010xe}.  
 
 As parametric resonance ceases to be efficient, there is a characteristic comoving length scale, $k_{\nl}$  where the fluctuations in the field are most nonlinear, namely the mode where the power spectrum peaks (see Figure \ref{fig:floquet}(b) and Figure \ref{fig:spectrum}). Following \cite{Amin:2010xe}, the number density of oscillons can be estimated as 
 \begin{equation}
 n_{\textrm{osc}}a^3\sim \left(\frac{k_{\nl}}{2\pi}\right)^3,
 \end{equation}
where the characteristic scale $k_\nl$ is given by
\begin{equation}
k_{\nl}\sim \left(\sqrt{\frac{2}{3}}\beta\right)^{-1/5}\z m
\end{equation}
with $\beta=\sqrt{\lambda}\z(\mpl/m)$. \footnote{Note that compared to the formula presented in \cite{Amin:2010xe}, there is an extra factor of $\sqrt{2/3}$ multiplying $\beta$ because we are starting with a different Hubble parameter $H_i$.} Thus
\begin{equation}
 n_{\textrm{osc}}a^3\sim \left(\sqrt{\frac{2}{3}}\beta\right)^{-3/5}\left(\frac{\lambda}{g}\frac{m}{2\pi}\right)^3.
 \label{eq:nosc3d}
 \end{equation}
Note that this is a rough estimate. Heuristically, we are merely counting the number of large peaks in the energy density. The linear analysis tells us how many such peaks we should expect in a given volume. 

    
\section{Numerical Simulations in 3 Dimensions}
\label{sec:ns3}
 %
    \begin{figure}[tbp] 
     \centering
     \includegraphics[width=3in]{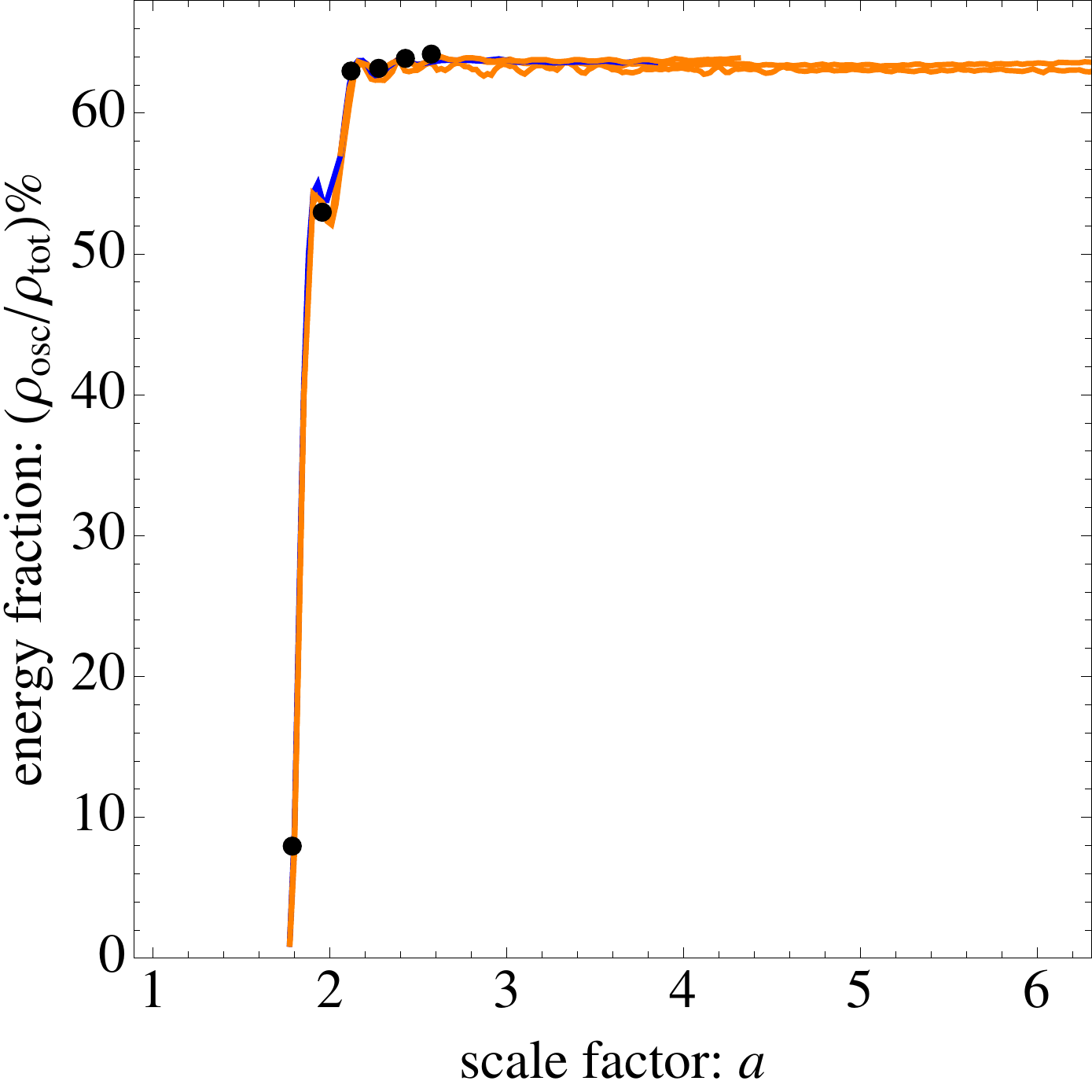} 
     \caption{The orange curves represent the evolution of the energy fraction in oscillons as we vary the initial boxsize $L=(200, 400, 600 m^{-1})$ with $N=256$. The blue curve shows the same quantity when $N=384$ with $L=400m^{-1}$. Finally the black points are based on  {\sc Defrost} output at $L=400 m^{-1}$ and $N=1024$. The agreement between the curves provides evidence for the robustness of our results. Note that this energy fraction is an underestimate since here we have defined the oscillon energy as the energy within a radius where the amplitude falls by $1/e$ of its core value. }
     \label{fig:LNCompare}
       \end{figure}
Numerically simulating the formation and evolution of oscillons in three dimensions requires the same tools and techniques developed during the analysis of parametric resonance and preheating, which greatly facilitates our work.  That said, oscillon dynamics present significant numerical challenges as the oscillons are relatively small and well separated in space, so we need to simulate a large volume with good spatial resolution. Moreover, we want to ensure that our results are cosmologically stable, and thus follow the evolution for several Hubble times ($H^{-1}\sim 10^3 m^{-1}$), which is a much larger interval than the natural oscillation period of the fields ($T\sim m^{-1}$).  

\subsection{Field Evolution Programs}
To ensure that our results are independent of our numerical algorithms, we performed simulations with a
variant of {\sc Defrost\/} \cite{Frolov:2008hy}, a finite differencing code, and {\sc PSpectRe}, which uses a pseudo-spectral scheme \cite{Easther:2010qz}.  For the cases we compared, both codes yield similar results and, unless otherwise specified, all results presented here were obtained with {\sc PSpectRe}.   We extended  {\sc Defrost\/} to include the sixth order scalar field potential and adapted it for use on an MPI-based cluster, running at simulation volumes of up to $1024^3$ points. {\sc PSpectRe} runs on multiple cores on a single shared-memory node, and these simulations were typically performed at $256^3$.  

The relative performance of both codes as a function of grid-size is discussed in  \cite{Easther:2010qz}. {\sc PSpectRe} typically reaches the ``continuum'' limit for smaller values of $N$ than a finite differencing code, offsetting the numerical overhead induced by the repeated Fourier transforms needed to construct the nonlinear terms in momentum space.  For both codes, simulations could require several days of wallclock time. At $1024^3$,  {\sc Defrost\/} parallelized efficiently  over $\sim 12$ nodes, or $\sim 100$ MPI threads. For some parameter values the thin, higher order resonance bands generate short-wavelength inhomogeneity after oscillon formation (see Section \ref{sec:spectrum}). In these cases, the relevant modes must be several times larger than the {\sc Defrost\/} grid resolution in order to ensure numerical fidelity.

We ran numerical simulations using both {\sc PSpectRe\/} and {\sc Defrost}  for a wide variety of free parameters, as well as for a range of choices of comoving boxsize $L$ and numerical resolution, measured by $N^3$ where $N^3$ is the number of points used by {\sc Defrost\/}, twice the number of momentum modes resolved by {\sc PSpectRe\/}. Figure \ref{fig:LNCompare} shows how the energy fraction in oscillons, time of emergence etc. is independent of intial box size $L$ and the spatial resolution $L/N$ as well as the code used.

\subsection{Oscillon Detection Algorithm}
 
Operationally, we define an  oscillon to be any contiguous region of space   whose energy density is at least $4$ times the mean density.  We detect oscillons by labeling each point over the threshold (on a position-space grid) with a unique, positive integer. We then iteratively set any pair of adjacent numbers  to the lower value, until no further changes are possible.  This creates a ``mask'', and a given oscillon consists of the set of grid points with the same numerical label. Once the oscillons are detected they can be trivially counted. 

For comparison with analytic results for the oscillon profiles, we define the volume of the oscillon to be the region for which the energy density is within $1/e$ of the peak density, and then compute the effective radius   $R = (3 V/4\pi)^{1/3}$. The total mass of the oscillon is given by the integral over the energy density within this volume. For steep-sided, flat-topped objects this definition captures a large fraction of the total energy density in the object,  but can exclude the shoulders of a  Gaussian oscillon.  Consequently, when we calculate the overall fraction of the energy in oscillons we will also consider the  region within $1/3e$ of the peak density.

In practice, we sometimes find non-oscillonic, isolated spikes that exceed  our threshold energy density. Rather than raising the threshold, we can  filter these objects by applying a cut on the energy of the spikes, as they are typically factor of 100 or more less massive than genuine oscillon configurations.  Conversely, while oscillons are typically well-separated at late times, interactions do occur, and this algorithm counts two interacting oscillons as a single object.  Finally, as a matter of convenience, we ran the oscillon detection code  on output files from {\sc PSpectRe} or {\sc Defrost}, rather than counting oscillons within these codes as they ran.

     \begin{figure}[tbp] 
     \centering
     \includegraphics[width=2.3 in]{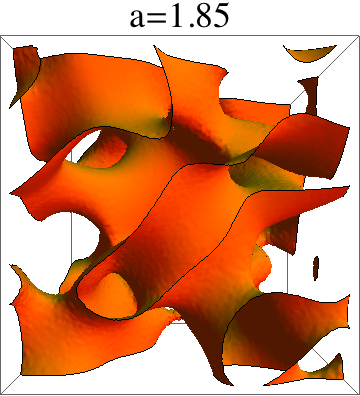} \includegraphics[width=2.3 in]{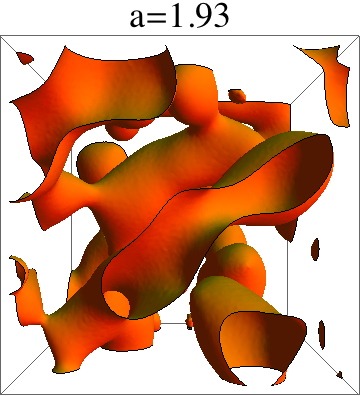} \\
        \includegraphics[width=2.3 in]{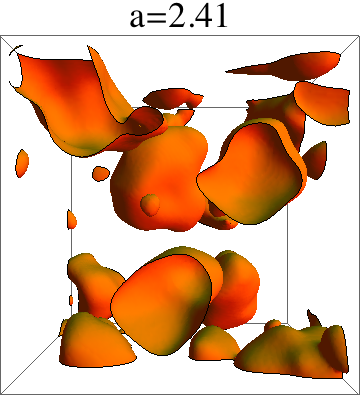} \includegraphics[width=2.3 in]{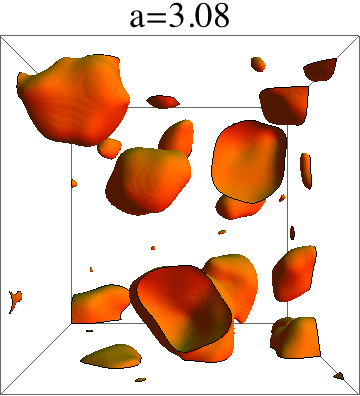} \\
        \includegraphics[width=2.3 in]{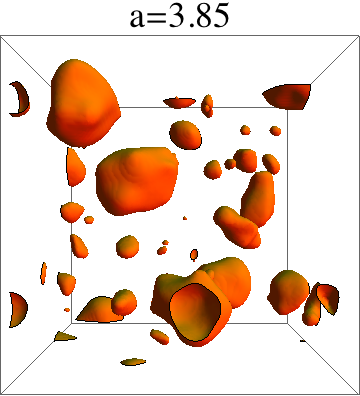} \includegraphics[width=2.3 in]{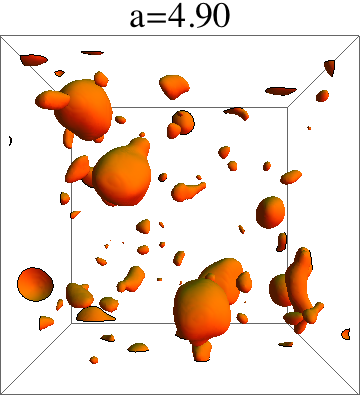}
       \caption{ Density contours ($\rho=\bar{\rho}$) for a $1/8$ sample of a simulation with fiducial parameters at $L = 200$ and $N = 256$. Frames are labelled by corresponding value of the scale factor. The large over-dense regions are oscillons; smaller regions are simply areas in which the density exceeds the average value, without being oscillonic. The frames show a fixed comoving volume.
         \label{fig:edgrid}}
  \end{figure}
  
      \begin{figure}[t] 
     \centering
     \includegraphics[width=2.87in]{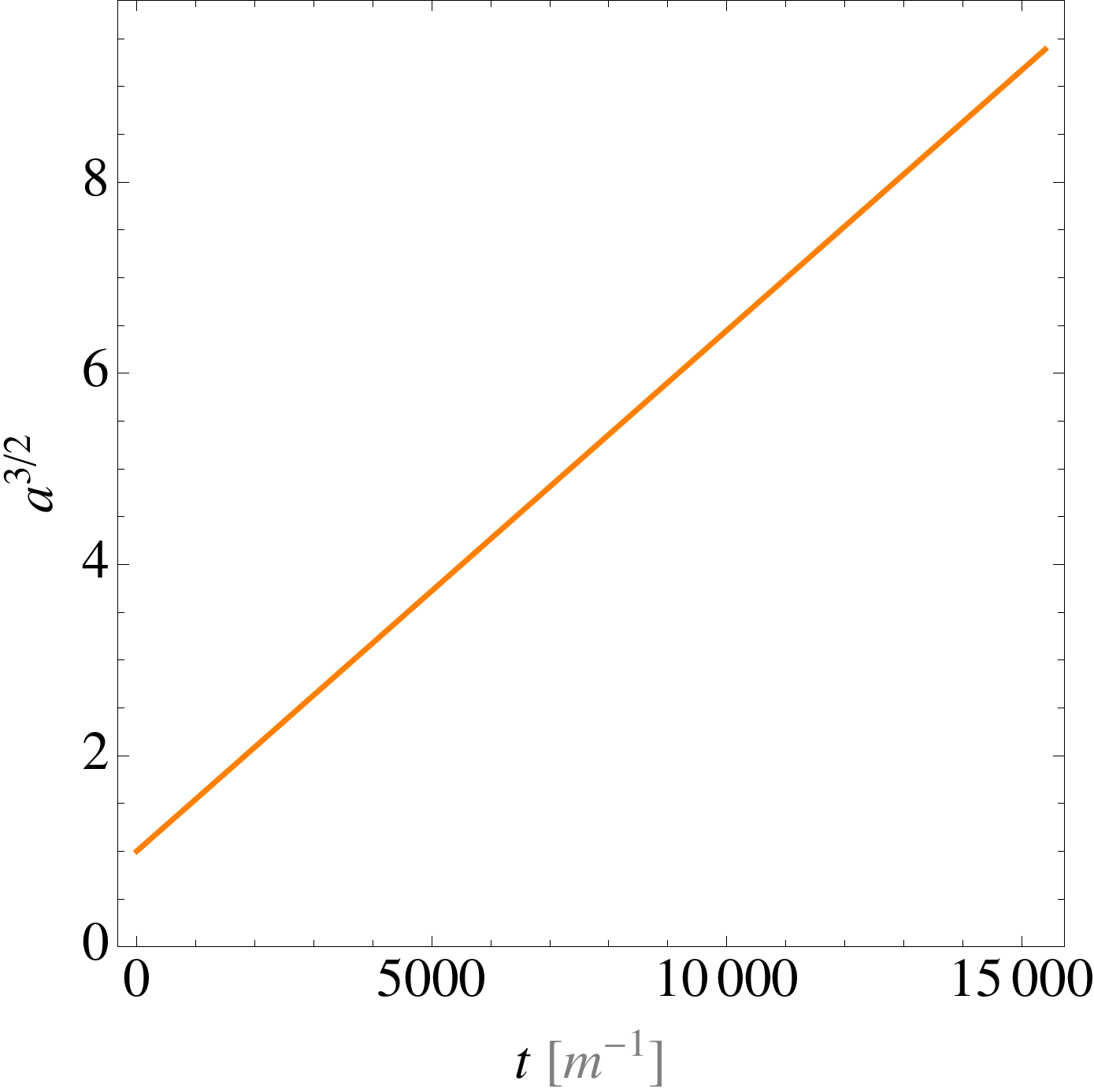} 
      \includegraphics[width=3in]{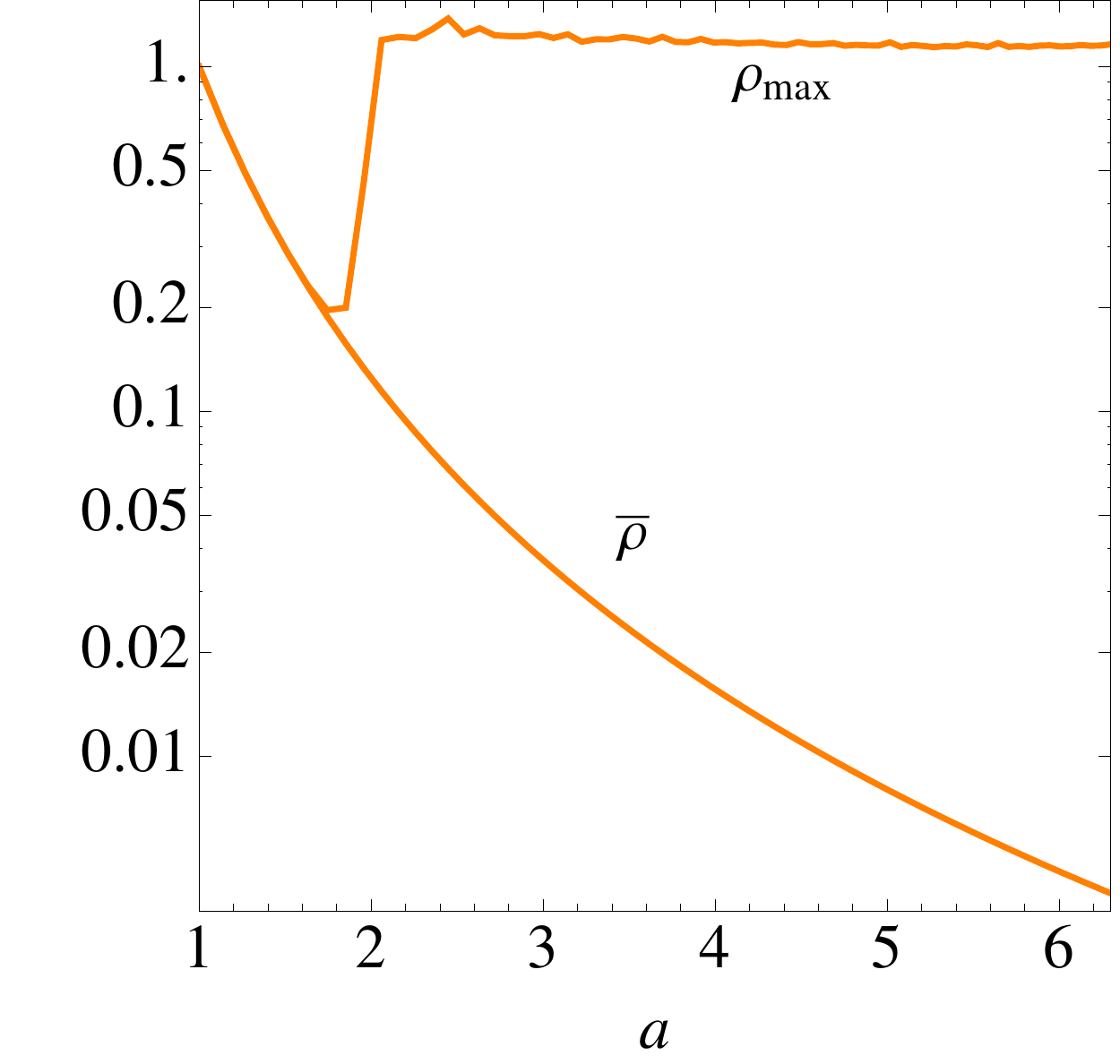} 
     \caption{The left panel shows  $a^{3/2}$ as a function of time $t$ for a ``fiducial'' simulation.  This is a linear function, consistent with the post-inflationary expansion mimicking that of a matter dominated universe. The right panel shows the maximum energy density in the simulation volume $\rho_{\textrm{max}}$, and the spatially-averaged energy density $\bar{\rho}$ as a function of  $a$. The former is roughly constant, suggesting that the maximum density of the oscillons is fixed. The density contrast associated with the oscillons grows as the average density is decreased by expansion. The initial phase of coherent oscillations during which $\rho_{\textrm{max}} \approx \bar{\rho}$ is decreasing is clearly visible in the second plot. The densities have been normalized to unity at $a=1$.}
     \label{fig:sf&contrast}
  \end{figure}

 \subsection{Qualitative Results}
 
 Unless otherwise stated,  we work with ``fiducial'' parameters $\lambda = 2.8125\times 10^{-6}$, $\z^2=0.1$ and $m=5\times 10^{-6}\mpl$.  Figure~\ref{fig:edgrid} depicts the spatial density profile seen in a typical simulation, at six different times. Surfaces are drawn at a overdensity of $1$, or an absolute density of 2, in units where $\bar{ \rho} \equiv 1$. Initially,  over-densities form on large comoving scales $\ll m^{-1}$, and collapse quickly into roughly tubular regions, which then fragment into the localized overdensities that we will identify as oscillons.  These overdense regions are well separated, and slowly moving with respect to the comoving background.   We see occasional interactions between these regions, but these are infrequent given their small spatial size (relative to the overall volume) and low velocity. Further, as the space expands, overdense regions occupy a progressively smaller comoving volume, consistent with analytic results suggesting that the  oscillons have approximately-fixed physical size. 

Figure~\ref{fig:sf&contrast} (left panel) shows that the expansion rate of the universe is entirely consistent with a pressureless fluid, for which $a(t) \propto t^{2/3}$, during the initial phase of coherent oscillations, parametric resonance, and after the onset of nonlinearity. This is contrast to many resonant scenarios \cite{Traschen:1990sw,Kofman:1994rk,Shtanov:1994ce,Kofman:1997yn, Podolsky:2005bw,GarciaBellido:2002aj, Felder:2001kt, Felder:2000hj, Tkachev:1998dc, Khlebnikov:1996mc,Greene:1998pb,Dufaux:2006ee} in which the universe is radiation-dominated or in some intermediate state \cite{Frolov:2008hy} following resonance.  This also confirms that while the potential contains terms beyond quadratic order to support the existence of oscillons, it is dominated by the $m^2 \varphi^2$ term  during the coherent oscillations phase.  Following resonance, Figure~\ref{fig:sf&contrast} (right panel) shows that the spatially-averaged density $\bar{\rho}$ falls rapidly, while the maximum density stays roughly constant. The maximum density is found at the ``core'' of the oscillons, and is expected to be constant as the universe expands.  Theoretically, we do expect oscillons to decay slowly, but at rates which are negligible on the time scale of our simulations.  To summarize, we have confirmed that  the resonant phase matches our expectations from the linearized theory \cite{Amin:2010xe}, as discussed in Section \ref{sec:icl},  while the ``global'' properties of the nonlinear phase are consistent with an oscillon dominated universe, with   localized overdensities which are a factor of 100  times larger than the average density.


We now seek to confirm that the overdensities we see are, in fact,   oscillons.   We show the relationship between energy density and width (volume) for  regions flagged as oscillons in the left panel of Figure \ref{fig:histhw}, and see an excellent match to the theoretical expectations of Amin and Shirokoff \cite{Amin:2010jq}. Note that at late times, almost all putative oscillons have $\rho> \rho_s$, consistent with the stability analysis in \cite{Amin:2010jq}.\footnote{This is in sharp contrast with the $1+1$ dimensional case where small amplitude oscillons are stable against long wavelength fluctuations.  As a result the left half of the corresponding ``U" shaped curve is equally populated \cite{Amin:2010jq,Amin:2010xe}.} An equatorial cross-section of a flat-topped oscillon and a Gaussian oscillon extracted from a simulation is shown in the right panel of Figure \ref{fig:flattops}. The white layer represents the single parameter analytic fit based on equation \eqref{eq:3dprofile}.  Clearly, a  Gaussian profile would not provide a good fit to the flat-topped oscillons. 

\subsection{Statistics, oscillon number and energy-density evolution}

\begin{figure}[tbp] 
     \centering
     \includegraphics[width=5.5in]{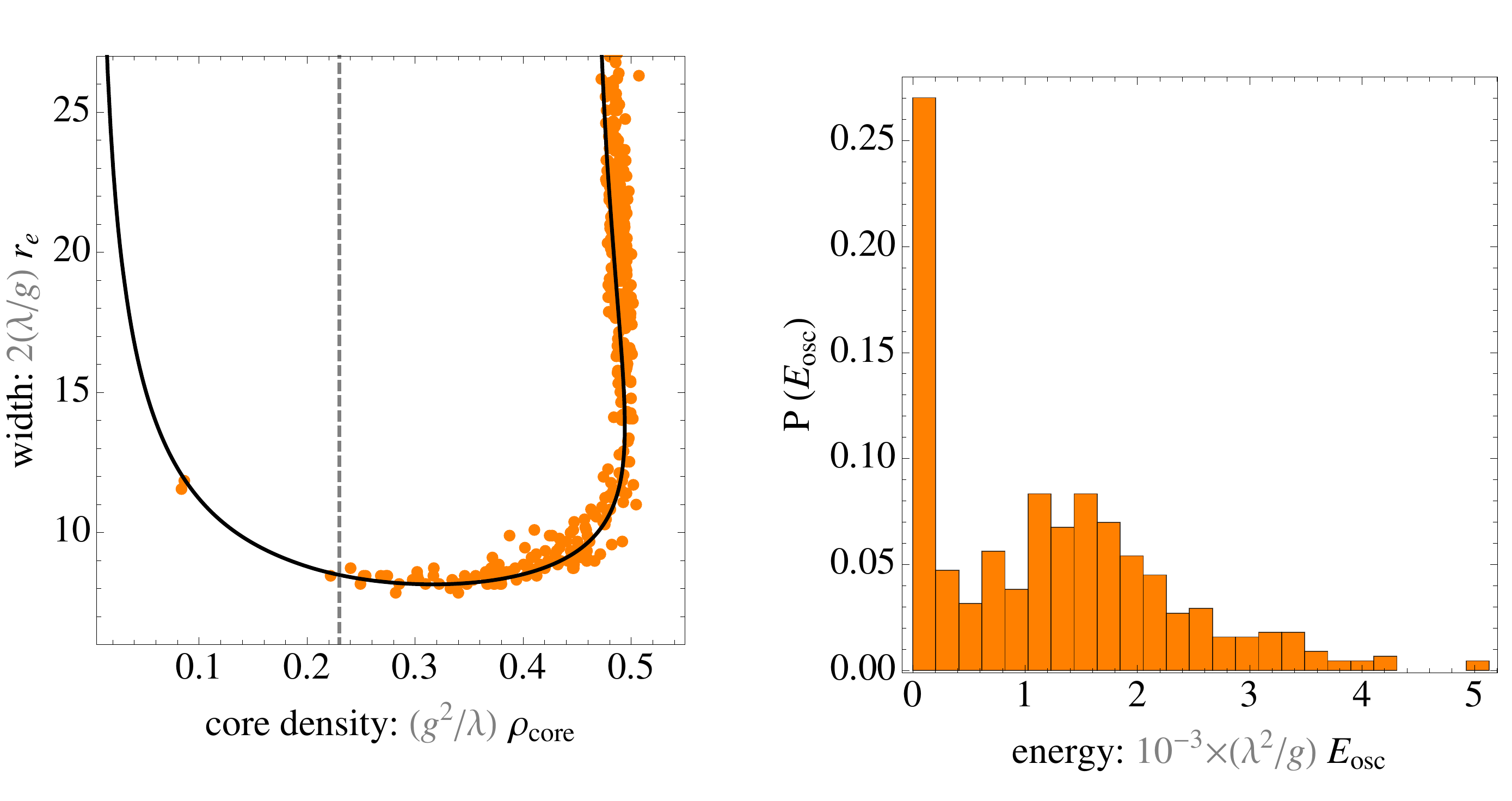} 
     \caption{The left panel  shows the non-monotonic relationship between the width and core amplitude of the oscillon energy densities, relative to the analytic prediction in \cite{Amin:2010jq} (solid black line). The width is taken as twice the radius where the energy density falls to $1/e$ of its core value. Orange points are overdensities flagged as oscillons in our simulations at $a\approx 5$. At large core amplitude we approach a flat-topped profile. The lack of oscillons at $(g^2/\lambda)\rho_{core}<(g^2/\lambda)\rho_{\alpha_s}\approx 0.23$ is consistent with the stability analysis of Amin and Shirokoff \cite{Amin:2010jq}. The right panel shows a histogram of oscillon energies with the width $r_e$. The parameters used were $\lambda\approx 2.8\times10^{-6}, \z^2=10^{-1}$ and  $m/\mpl=5\times 10^{-6}$.  
All variables are expressed in units of appropriate powers of mass, $m$.}
     \label{fig:histhw}
  \end{figure}

Figure \ref{fig:histhw} (right panel) shows a histogram of oscillon energies at late times, when $a_f\approx5$. Qualitatively, we see large number of small oscillons and a weaker peak in the distribution at higher energies. This interpretation is somewhat subjective, but resembles the results from the one dimensional case, which had two clear ``generations" of oscillons \cite{Amin:2010xe}.   Oscillons in the right peak (more massive) form first, and the number density of {\it first generation} oscillons matches well with the analytic estimate of  equation \eqref{eq:nosc3d} \cite{Amin:2010xe}.  However, we note that this distinction between ``generations" cannot always be made cleanly, especially for large $\beta$ values. For this paper, we do not distinguish between the two generations.

\begin{figure}[tbp] 
     \centering
     \includegraphics[width=5.5in]{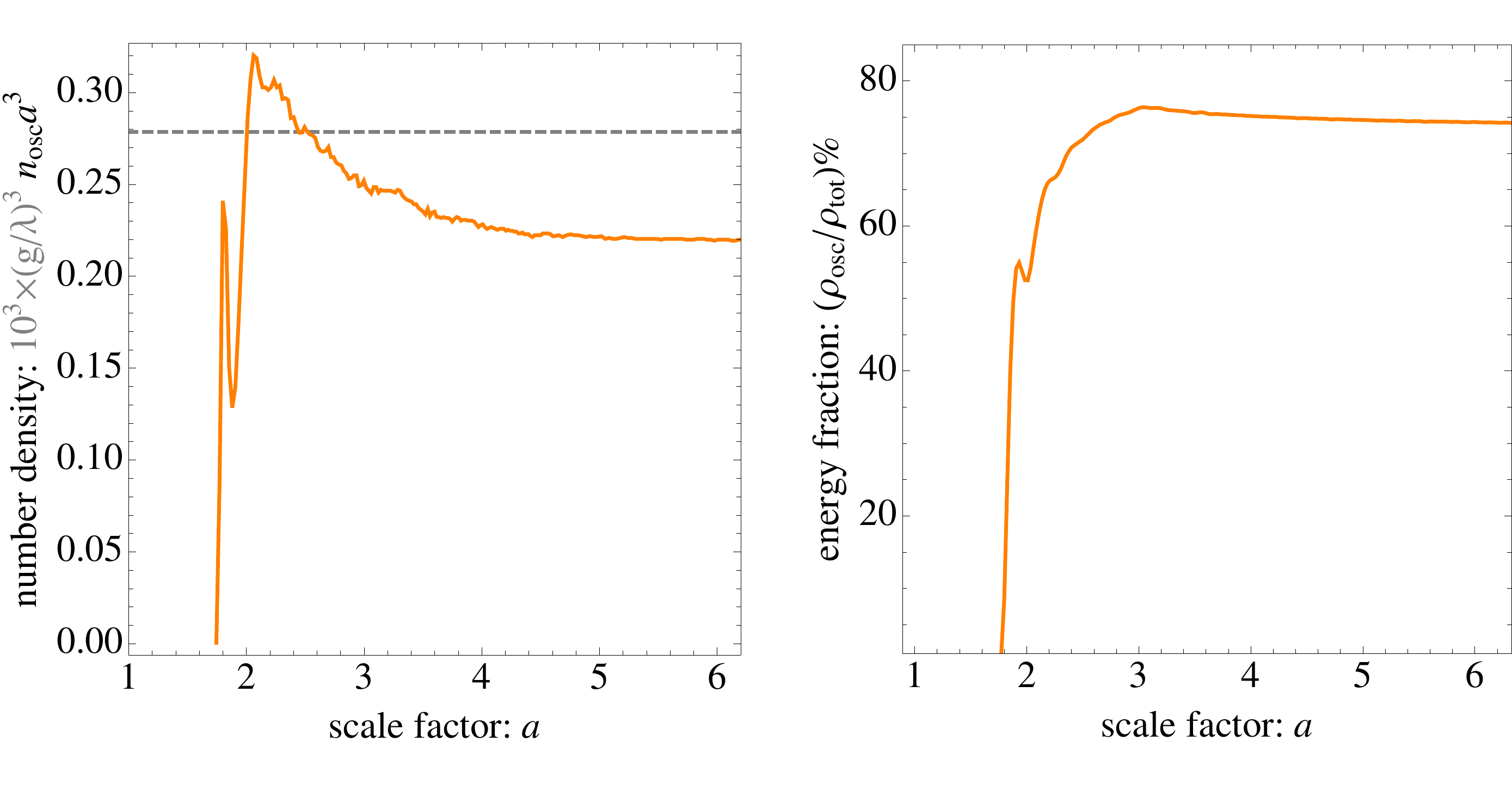} 
     \caption{The left panel shows the number density of oscillons as a function of scale factor. Oscillons emerge at $a_{\nl}\sim 1.5$. Smaller oscillons are transient, leaving  a roughly constant (comoving) number density of large-amplitude oscillons. The dashed line is the analytic stability estimate from Amin \cite{Amin:2010xe}. The energy density in oscillons  is shown on the right. The fraction of the total energy density in oscillons, defined here as regions in which $\rho>\rho_{\textrm{core}}/3e$ for individual oscillons, is shown in the right panel.   The parameters used were $\lambda\approx 2.8\times10^{-6}$ (or $\beta=10^2$), $\z^2=0.1$ and $m/\mpl=5\times 10^{-6}$. Number density is expressed in units of $m^3$. }
     \label{fig:enosca}
  \end{figure}

Figure \ref{fig:enosca} (right) shows the comoving number density of oscillons as a function of time.  Oscillons are produced at $a_{\nl}\sim 1.7$, and the number density peaks there. As noted above, small oscillons are unstable in three dimensions, but  the  number density stabilizes within a Hubble time. The remaining large oscillons have a fixed physical size and thus shrink, relative to the comoving simulation volume, and their typical separation scales with the expansion. The dashed line is the analytic estimate based on equation \eqref{eq:nosc3d}  \cite{Amin:2010xe}. The fraction of energy density in oscillons is shown in Figure \ref{fig:enosca}(left).  Note that the fractional energy computed here underestimates the total energy associated with the oscillons. At late times in our simulations, the peak density can be at least 100 times larger than the average. Even if we include points whose energy is within $1/3e$ of the maximum, we still ignore points for for which density contrast is of order 10.  Qualitatively, the vast majority of the energy density is accounted for by the oscillons. 

\begin{figure}[tbp] 
     \centering
     \includegraphics[width=3in]{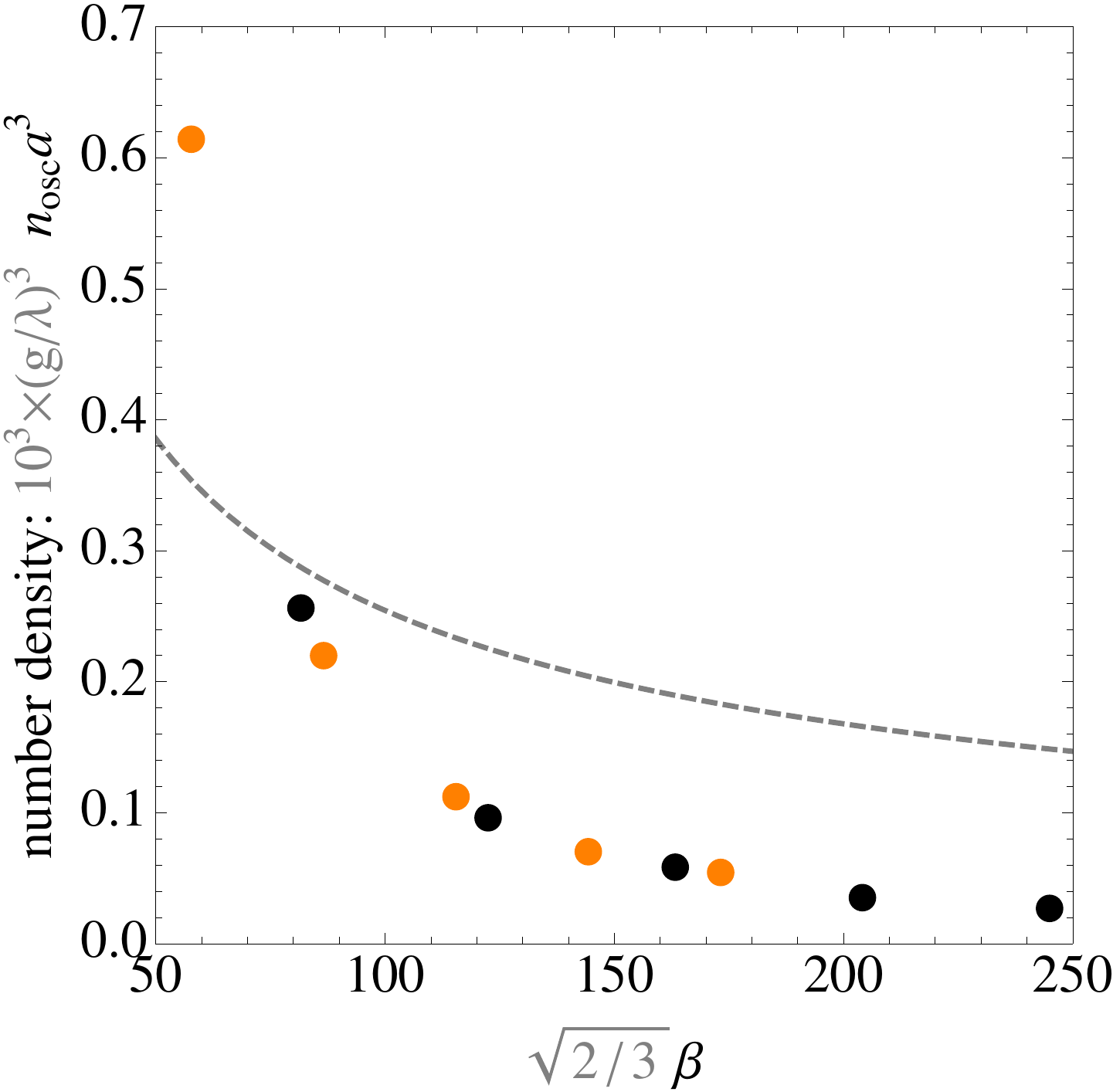} 
     \caption{The above Figure shows the variation of the co-moving number density with $\beta=\lambda^{3/2}g(\mpl/m)$ which characterizes the ratio between the maximum growth rate of fluctuations in flat space and the Hubble parameter. The dashed curve is the analytic estimate provided by Amin \cite{Amin:2010xe}. The orange points are for $(\lambda/g)^2=0.1$ and the black ones are for $(\lambda/g)^2=0.2$. The dependence on $\z$ is accurately captured by the analytic estimate, which results in the orange and black points lying on the same curve. However the dependence on the relative growth rate of fluctuations, $\beta$ is stronger than in the analytic estimate. The deviation from the analytic estimates can be attributed to oscillon-oscillon collisions at small Hubble values (large $\beta$). }
     \label{fig:nlfirst3d}
  \end{figure}

  We   further check our expectations by  varying $\lambda$ and $g$, and verifying that the numerical results reproduce the scalings seen in  semianalytic estimates of the co-moving number density of large oscillons  \cite{Amin:2010xe}. We set $L=400m^{-1}$ and $N=256$ and express the number density as a function of $\beta=\sqrt{\lambda}\z(\mpl/m)$, since $\beta$ roughly characterizes the strength of resonance in the linear regime. 
Figure \ref{fig:nlfirst3d}(b) shows the final number density of oscillons at late times, after the oscillon number density has stabilized. We show the results as a function of  $\beta$ for $\z^2=0.1$ (orange) and $\z^2=0.2$ (black), while $\lambda$ runs from $1.25\times 10^{-6}$ to $1.125\times 10^{-5}$. The dashed curve  is the analytic estimate derived by Amin in \cite{Amin:2010xe}. Remarkably, despite the non-linear dynamics of oscillon formation, the analytic estimate is good to within a factor of a  a few and correctly captures the parameter dependence.  In particular, the variation with $\z$ is almost exactly described by the scaling of the vertical axis. 

The analytic result over-estimates the number of oscillons at large $\beta$, where the expansion rate is slow. Consequently, one can infer that oscillon-oscillon interactions are more common at formation, which appears to lead to mergers and fewer, larger oscillons and this process was not part of our  estimate.  The outcome of a collision depends on the phase, amplitude, velocity as well as shape of the oscillons (see \cite{Hindmarsh:2007jb}). The collisions can be almost completely elastic ones  or highly inelastic ones, though we are almost always left with one or more oscillons. Details of oscillon-oscillons collisions and fragmentation will be considered in future work.  

\begin{figure}[tbp] 
     \centering
     \includegraphics[width=5.5in]{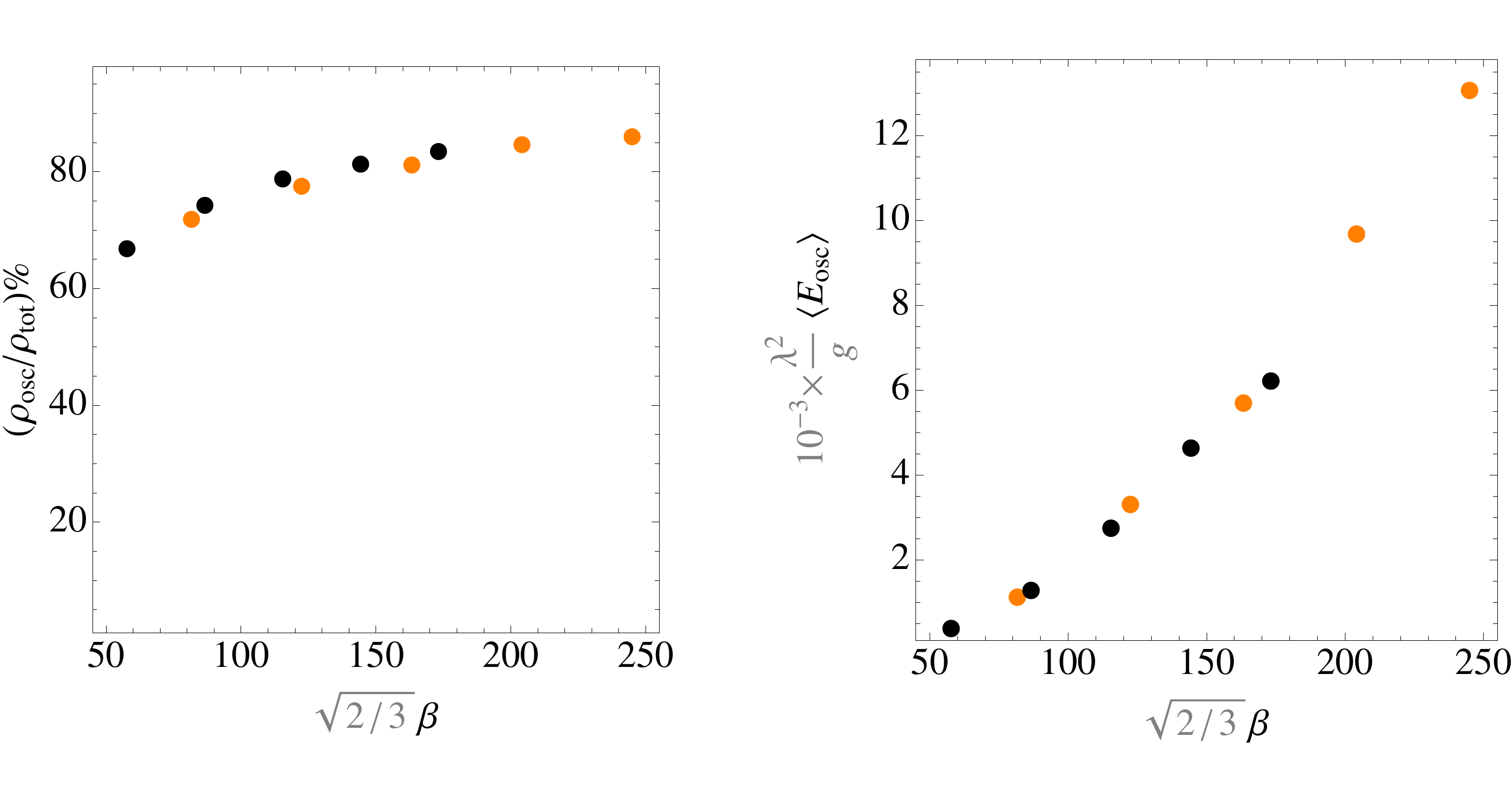} 
     \caption{The left panel shows the percentage of energy density in oscillons at the end of our simulations for different values of the relative growth rate of fluctuations: $\beta=\sqrt{\lambda}\z(\mpl/m)$. The black and orange points correspond to simulations with parameters $\z^2=0.2$ and $0.1$ respectively. The fraction of energy density in oscillons decreases at low $\beta$ (large $H$) because resonance is not efficient enough to significantly amplify initial fluctuations. The right panel shows the mean energy per oscillon. Note that it increases with increasing $\beta$ (decreasing $H$). This is consistent with the idea that at large $\beta$ oscillons form from longer wavelength fluctuations, thus yielding fewer, larger oscillons. The orange and black points are aligned as expected, because of the scaling with $\z$ on the vertical axes. Energy is expressed in units of the mass, $m$. }
     \label{fig:efracE}
  \end{figure}

As noted previously,  the universe is effectively oscillon-dominated after nonlinearity sets in. In Figure \ref{fig:efracE} (left), we quantify this statement. At very small $\beta$, we cannot generate oscillons efficiently, so the oscillon contribution initially increases with $\beta$, but eventually saturates. In Figure \ref{fig:efracE} (right), we show the average energy per oscillon as a function of $\beta$. As $\beta$ increases ($H$ decreases), we get fewer, more massive oscillons. This is consistent with the shape of the resonance band seen in Figure \ref{fig:floquet}(a): for small $H$ the longer wavelength modes have a chance to get amplified before the ``sweet spot" of the Floquet band is traversed by $k\sim \z m$ modes. Further, as discussed above, small $H$ also allows for a larger number of interactions between oscillons, possibly leading to mergers and hence bigger oscillons.


\subsection{Spectrum of linear and non-linear fluctuations}
\label{sec:spectrum}
    \begin{figure}[tbp] 
     \centering
     \includegraphics[width=5.5in]{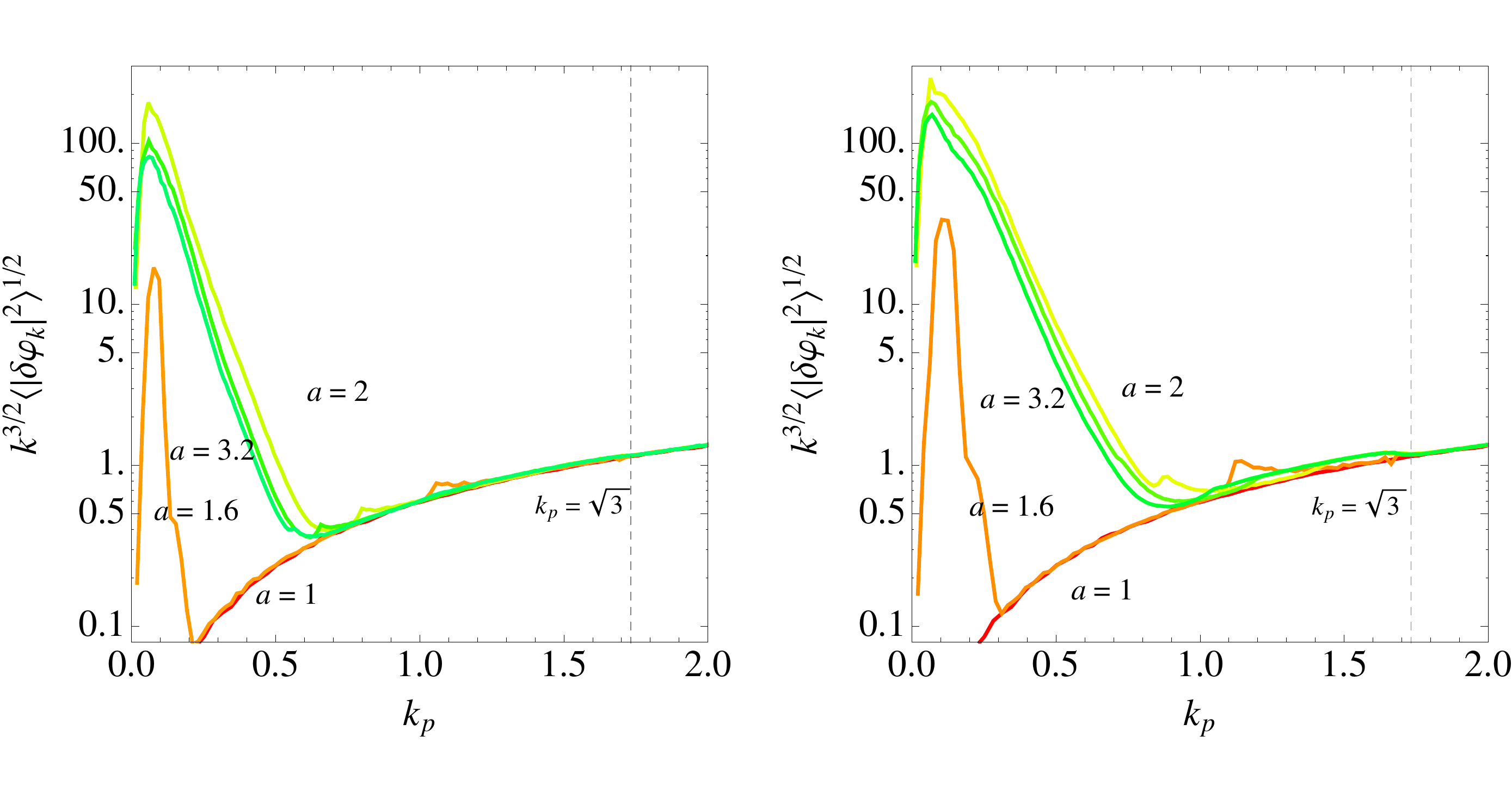} 
     \caption{The figures show the evolution of the spectrum of field fluctuations for $\lambda=2.8125\times10^{-6}$ and $\z^2=0.1$ (left panel) and $\z^2=0.2$ (right panel). The orange and red curves show the linear evolution of fluctuations through resonance. This broad band is responsible for oscillon production. The small bump seen at physical momentum $k_{\textrm{p}}\sim m$ is due to higher-order resonance bands excited by the homogeneous field initially as well as by very wide oscillons at late times. The power increase in power at $k_{\textrm{p}}\sim \sqrt{3}m$ is clearly visible in the right panel. }
     \label{fig:spectrum}
       \end{figure}

As we have seen the initial, linear growth of fluctuations is accurately and conveniently described in Fourier space, whereas oscillons find a natural description in position space. Nevertheless, it is instructive to look at the spectrum of fluctuations through the entire evolution.        
Figure \ref{fig:spectrum} shows the spectrum of fluctuations for the parameters $\lambda\approx2.8\times10^{-6}$ and $\z^2=0.1$ (left panel) and $\z^2=0.2$ (right panel). The orange and red curves represent the spectrum in the linear regime, whereas the yellow and green ones are deep in the non-linear regime. The biggest feature in the spectrum is from the band that produces oscillons initially and at late time shows the Fourier content of spatially localized oscillons.

There is an additional, tiny bump in the spectrum at the {\it physical} momentum $k_{\textrm{p}}\sim m$.  A narrow,  higher-order resonance band at $k_{\textrm{p}}\sim\sqrt{3}m$ generates this bump during the initial phase of coherent oscillations. That power is redshifted and this small bump moves to the left. At late times, long after resonance due to the homogenous field has ended, very wide oscillons regenerate power at $k_{\textrm{p}}\sim\sqrt{3}m$, as they  effectively behave as a homogeneous field. This can be seen in the slight increase in power below $k_{\textrm{p}}\sim\sqrt{3}m$ (left panel). This Floquet instability at $k_{\textrm{p}}\sim \sqrt{3}m$ \cite{Amin:2010jq, Hertzberg:2010yz} can potentially destabilize the oscillons and drain their energy. However, these fluctuations are catastrophic for the oscillon only if fluctuations cannot escape the oscillon before being significantly amplified. This yields a relation between the rate of growth of fluctuations and the width of the oscillons. We refer the reader to  \cite{Hertzberg:2010yz} for further details. As seen in Figure \ref{fig:spectrum},  we see this effect in $k_{\textrm{p}}\sim \sqrt{3}m$ fluctuations clearly for case $\z^2=0.2$ (right panel) but not for the case $\z^2=0.1$ (left panel) at late times. This is expected since the Floquet exponent  for the $k\sim\sqrt{3}m$ instability band scales as $\propto \z^2$. We note that for many runs we cannot resolve the $k_{\textrm{p}}\sim \sqrt{3}m$ instability towards the end of the simulation because of the degraded resolution from the expanding background. For some sample higher resolution runs, we did not see a significant change in the number density of oscillons. Nevertheless, the effect of this instability needs to be investigated further. 

 \section{Discussion}
\label{sec:disc}
In this paper we have numerically investigated the post-inflationary emergence of oscillons in a class of single-field potentials   in $3+1$ dimensions.  Beginning our simulations with a homogeneous inflaton condensate and the zero point fluctuations of the inflaton, resonance-driven   inflaton fragmentation leads to the copious production of oscillons.  In these scenarios, inflation is followed by a period of oscillon domination, during which the dominant component of the universe is localized ``blobs'' of scalar field matter.  The universe is pressureless, leading to an effective matter-dominated phase {\em following\/} parametric resonance.     The number density  and individual characteristics of oscillons closely match the theoretical predictions of \cite{Amin:2010jq} and \cite{Amin:2010xe}. 

This paper forms part of a wider investigation of the  existence and consequences of oscillon-like field configurations in the very early universe. This work is based on a potential which is symmetric  as $\varphi \rightarrow -\varphi$, truncated at the third nontrivial term in the Taylor expansion which is taken to be large, in order to make contact with the analytic discussion of \cite{Amin:2010jq, Amin:2010xe}.  Our next step in this program will be to relax these conditions, and demonstrate the existence of persistent, oscillon-like configurations for realistic inflationary potentials, and to explore the cosmological consequences of a post-inflationary, oscillon-dominated phase.  Looking further ahead, we wish to understand the impact of the higher-order resonance bands on oscillon dynamics, account for  the effects of couplings to additional fields, consider the impact of gravitational effects on the formation and evolution of isolated oscillons, and the interactions between oscillons.

 \section*{Acknowledgements}
RE and HF are partially supported by the United States Department of Energy (DE-FG02-92ER-40704) and the National Science Foundation (CAREER-PHY-0747868).  HF is supported, in part, by the United States Department of Energy Computational Science Graduate Fellowship, provided under grant DE-FG02-97ER25308. MA is supported by a Pappalardo fellowship at MIT. This work was supported in part by the facilities and staff of the Yale University Faculty of Arts and Sciences High Performance Computing Center. We particularly thank Brian Dobbins for facilitating the numerical simulations and Daniel Kapec for providing us with a C-based  implementation of our oscillon detection algorithm.  This work made significant use of Andrei Frolov's code {\sc Defrost}, and we are grateful him for making this package publicly available. We thank Eddie Farhi,  Raphael Flauger, Alan Guth, Mark Hertzberg, Mark Mezei,  Ruben Rosales, Evangelos Sfakianakis and David Shirokoff for helpful discussions.


\begin{thebibliography}{77}

\bibitem{Amin:2010jq}
  M.~A.~Amin and D.~Shirokoff,
  arXiv:1002.3380 [astro-ph.CO].
  
 
  
\bibitem{Amin:2010xe}
  M.~A.~Amin,
  arXiv:1006.3075 [astro-ph.CO].
  

\bibitem{Bogolyubsky:1976yu}
  I.~L.~Bogolyubsky and V.~G.~Makhankov,
  Pisma Zh.\ Eksp.\ Teor.\ Fiz.\  {\bf 24}, 15 (1976).
  
\bibitem{Gleiser:1993pt}
  M.~Gleiser,
  Phys.\ Rev.\  D {\bf 49}, 2978 (1994)
  [arXiv:hep-ph/9308279].

\bibitem{Honda:2001xg}
  E.~P.~Honda and M.~W.~Choptuik,
  Phys.\ Rev.\  D {\bf 65}, 084037 (2002)
  [arXiv:hep-ph/0110065].
  
\bibitem{Adib:2002ff}
  A.~B.~Adib, M.~Gleiser and C.~A.~S.~Almeida,
  Phys.\ Rev.\  D {\bf 66}, 085011 (2002)
  [arXiv:hep-th/0203072].

\bibitem{Farhi:2005rz}
  E.~Farhi, N.~Graham, V.~Khemani, R.~Markov and R.~Rosales,
  Phys.\ Rev.\  D {\bf 72}, 101701 (2005)
  [arXiv:hep-th/0505273].

  



\bibitem{Gleiser:2006te}
  M.~Gleiser,
  Int.\ J.\ Mod.\ Phys.\  D {\bf 16}, 219 (2007)
  [arXiv:hep-th/0602187].
  
\bibitem{Graham:2006xs}
  N.~Graham and N.~Stamatopoulos,
  Phys.\ Lett.\  B {\bf 639}, 541 (2006)
  [arXiv:hep-th/0604134].


\bibitem{Hindmarsh:2006ur}
  M.~Hindmarsh and P.~Salmi,
  Phys.\ Rev.\  D {\bf 74}, 105005 (2006)
  [arXiv:hep-th/0606016].
  
\bibitem{Fodor:2006zs}
  G.~Fodor, P.~Forgacs, P.~Grandclement and I.~Racz,
  Phys.\ Rev.\  D {\bf 74}, 124003 (2006)
  [arXiv:hep-th/0609023].



\bibitem{Saffin:2006yk}
  P.~M.~Saffin and A.~Tranberg,
  JHEP {\bf 0701}, 030 (2007)
  [arXiv:hep-th/0610191].


\bibitem{Graham:2006vy}
  N.~Graham,
  Phys.\ Rev.\ Lett.\  {\bf 98}, 101801 (2007)
  [Erratum-ibid.\  {\bf 98}, 189904 (2007)]
  [arXiv:hep-th/0610267].


\bibitem{Arodz:2007jh}
  H.~Arodz, P.~Klimas and T.~Tyranowski,
  Phys.\ Rev.\  D {\bf 77}, 047701 (2008)
  [arXiv:0710.2244 [hep-th]].


\bibitem{Hindmarsh:2007jb}
  M.~Hindmarsh and P.~Salmi,
  Phys.\ Rev.\  D {\bf 77}, 105025 (2008)
  [arXiv:0712.0614 [hep-th]].


\bibitem{Fodor:2008es}
  G.~Fodor, P.~Forgacs, Z.~Horvath and A.~Lukacs,
  Phys.\ Rev.\  D {\bf 78}, 025003 (2008)
  [arXiv:0802.3525 [hep-th]].


\bibitem{Gleiser:2008ty}
  M.~Gleiser and D.~Sicilia,
  Phys.\ Rev.\ Lett.\  {\bf 101}, 011602 (2008)
  [arXiv:0804.0791 [hep-th]].



\bibitem{Fodor:2008du}
  G.~Fodor, P.~Forgacs, Z.~Horvath and M.~Mezei,
  Phys.\ Rev.\  D {\bf 79}, 065002 (2009)
  [arXiv:0812.1919 [hep-th]].


\bibitem{Fodor:2009xw}
  G.~Fodor, P.~Forgacs, Z.~Horvath and M.~Mezei,
  JHEP {\bf 0908}, 106 (2009)
  [arXiv:0906.4160 [hep-th]].
  

  
\bibitem{Gleiser:2009ys}
  M.~Gleiser and D.~Sicilia,
  Phys.\ Rev.\  D {\bf 80}, 125037 (2009)
  [arXiv:0910.5922 [hep-th]].



\bibitem{Gleiser:2007te}
  M.~Gleiser and J.~Thorarinson,
  Phys.\ Rev.\  D {\bf 76}, 041701 (2007)
  [arXiv:hep-th/0701294].
  
\bibitem{Hertzberg:2010yz}
  M.~P.~Hertzberg,
  arXiv:1003.3459 [hep-th].


\bibitem{Gleiser:2003uu}
  M.~Gleiser and R.~C.~Howell,
  Phys.\ Rev.\  E {\bf 68}, 065203 (2003)
  [arXiv:cond-mat/0310157].
  
\bibitem{Copeland:1995fq}
  E.~J.~Copeland, M.~Gleiser and H.~R.~Muller,
  Phys.\ Rev.\  D {\bf 52}, 1920 (1995)
  [arXiv:hep-ph/9503217].
  
\bibitem{Farhi:2007wj}
  E.~Farhi, N.~Graham, A.~H.~Guth, N.~Iqbal, R.~R.~Rosales and N.~Stamatopoulos,
  Phys.\ Rev.\  D {\bf 77}, 085019 (2008)
  [arXiv:0712.3034 [hep-th]].


\bibitem{Dymnikova:2000dy}
  I.~Dymnikova, L.~Koziel, M.~Khlopov and S.~Rubin,
  Grav.\ Cosmol.\  {\bf 6}, 311 (2000)
  [arXiv:hep-th/0010120].




  
  

\bibitem{McDonald:2001iv}
  J.~McDonald,
  Phys.\ Rev.\  D {\bf 66}, 043525 (2002)
  [arXiv:hep-ph/0105235].
  
  
\bibitem{Broadhead:2005hn}
  M.~Broadhead and J.~McDonald,
  Phys.\ Rev.\  D {\bf 72}, 043519 (2005)
  [arXiv:hep-ph/0503081].
  
\bibitem{Copeland:2002ku}
  E.~J.~Copeland, S.~Pascoli and A.~Rajantie,
  Phys.\ Rev.\  D {\bf 65}, 103517 (2002)
  [arXiv:hep-ph/0202031].

\bibitem{Kasuya:2000wx}
  S.~Kasuya and M.~Kawasaki,
  Phys.\ Rev.\  D {\bf 62}, 023512 (2000)
  [arXiv:hep-ph/0002285].




  
\bibitem{Traschen:1990sw}
  J.~H.~Traschen and R.~H.~Brandenberger,
  Phys.\ Rev.\  D {\bf 42}, 2491 (1990).

  
\bibitem{Kofman:1994rk}
  L.~Kofman, A.~D.~Linde and A.~A.~Starobinsky,
  Phys.\ Rev.\ Lett.\  {\bf 73}, 3195 (1994)
  [arXiv:hep-th/9405187].
  
\bibitem{Shtanov:1994ce}
  Y.~Shtanov, J.~H.~Traschen and R.~H.~Brandenberger,
  Phys.\ Rev.\  D {\bf 51}, 5438 (1995)
  [arXiv:hep-ph/9407247].
  
\bibitem{Kofman:1997yn}
  L.~Kofman, A.~D.~Linde and A.~A.~Starobinsky,
  Phys.\ Rev.\  D {\bf 56}, 3258 (1997)
  [arXiv:hep-ph/9704452].
  
\bibitem{Podolsky:2005bw}
  D.~I.~Podolsky, G.~N.~Felder, L.~Kofman and M.~Peloso,
  Phys.\ Rev.\  D {\bf 73}, 023501 (2006)
  [arXiv:hep-ph/0507096].
  
\bibitem{GarciaBellido:2002aj}
  J.~Garcia-Bellido, M.~Garcia Perez and A.~Gonzalez-Arroyo,
  Phys.\ Rev.\  D {\bf 67}, 103501 (2003)
  [arXiv:hep-ph/0208228].
  
\bibitem{Felder:2001kt}
  G.~N.~Felder, L.~Kofman and A.~D.~Linde,
  Phys.\ Rev.\  D {\bf 64}, 123517 (2001)
  [arXiv:hep-th/0106179].
  
\bibitem{Felder:2000hj}
  G.~N.~Felder, J.~Garcia-Bellido, P.~B.~Greene, L.~Kofman, A.~D.~Linde and I.~Tkachev,
  Phys.\ Rev.\ Lett.\  {\bf 87}, 011601 (2001)
  [arXiv:hep-ph/0012142].

\bibitem{Tkachev:1998dc}
  I.~Tkachev, S.~Khlebnikov, L.~Kofman and A.~D.~Linde,
  Phys.\ Lett.\  B {\bf 440}, 262 (1998)
  [arXiv:hep-ph/9805209].
  
\bibitem{Khlebnikov:1996mc}
  S.~Y.~Khlebnikov and I.~I.~Tkachev,
  Phys.\ Rev.\ Lett.\  {\bf 77}, 219 (1996)
  [arXiv:hep-ph/9603378].

  \bibitem{Greene:1998pb}
  P.~B.~Greene, L.~Kofman and A.~A.~Starobinsky,
  Nucl.\ Phys.\  B {\bf 543}, 423 (1999)
  [arXiv:hep-ph/9808477].
  
\bibitem{Dufaux:2006ee}
  J.~F.~Dufaux, G.~N.~Felder, L.~Kofman, M.~Peloso and D.~Podolsky,
  JCAP {\bf 0607}, 006 (2006)
  [arXiv:hep-ph/0602144].

  
\bibitem{Segur:1987mg}
  H.~Segur and M.~D.~Kruskal,
  Phys.\ Rev.\ Lett.\  {\bf 58} (1987) 747.
  
\bibitem{Fodor:2009kf}
  G.~Fodor, P.~Forgacs, Z.~Horvath and M.~Mezei,
  Phys.\ Lett.\  B {\bf 674} (2009) 319
  [arXiv:0903.0953 [hep-th]].
  
\bibitem{Alpher:1948}
R.~A.~Alpher, H.~Bethe and G.~Gamow,
 Phys.\ Rev. {\bf 73}, 803 (1948).

  
\bibitem{Wagoner:1966pv}
  R.~V.~Wagoner, W.~A.~Fowler and F.~Hoyle,
  Astrophys.\ J.\  {\bf 148}, 3 (1967).
  
\bibitem{Khlopov:2008qy}
  M.~Y.~Khlopov,
  Res.\ Astron.\ Astrophys.\  {\bf 10}, 495 (2010)
  [arXiv:0801.0116 [astro-ph]].


  
  
  

 
  

  
%
%
%
%

%


\bibitem{Khlebnikov:1997di}
  S.~Y.~Khlebnikov and I.~I.~Tkachev,
  Phys.\ Rev.\  D {\bf 56}, 653 (1997)
  [arXiv:hep-ph/9701423].
  
\bibitem{Easther:2006gt}
  R.~Easther and E.~A.~Lim,
  JCAP {\bf 0604}, 010 (2006)
  [arXiv:astro-ph/0601617].
  
\bibitem{Dufaux:2007pt}
  J.~F.~Dufaux, A.~Bergman, G.~N.~Felder, L.~Kofman and J.~P.~Uzan,
  Phys.\ Rev.\  D {\bf 76}, 123517 (2007)
  [arXiv:0707.0875 [astro-ph]].


\bibitem{Easther:2006vd}
  R.~Easther, J.~T.~Giblin and E.~A.~Lim,
  Phys.\ Rev.\ Lett.\  {\bf 99}, 221301 (2007)
  [arXiv:astro-ph/0612294].
  
\bibitem{Easther:2007vj}
  R.~Easther, J.~T.~Giblin and E.~A.~Lim,
  Phys.\ Rev.\  D {\bf 77}, 103519 (2008)
  [arXiv:0712.2991 [astro-ph]].

  
\bibitem{GarciaBellido:2007af}
  J.~Garcia-Bellido, D.~G.~Figueroa and A.~Sastre,
  Phys.\ Rev.\  D {\bf 77}, 043517 (2008)
  [arXiv:0707.0839 [hep-ph]].

  \bibitem{Giblin:2010sp}
  J.~T.~Giblin, L.~R.~Price and X.~Siemens,
  arXiv:1006.0935 [astro-ph.CO].

  

  
  
%

%

 
  
%

  
  
  

  
  
  
\bibitem{Coleman:1985ki}
  S.~R.~Coleman,
  Nucl.\ Phys.\  B {\bf 262}, 263 (1985)
  [Erratum-ibid.\  B {\bf 269}, 744 (1986)].
  
\bibitem{Lee:1991ax}
  T.~D.~Lee and Y.~Pang,
  Phys.\ Rept.\  {\bf 221}, 251 (1992).
  
\bibitem{Enqvist:2002rj}
  K.~Enqvist, S.~Kasuya and A.~Mazumdar,
  Phys.\ Rev.\ Lett.\  {\bf 89}, 091301 (2002)
  [arXiv:hep-ph/0204270].
  
\bibitem{Enqvist:2002si}
  K.~Enqvist, S.~Kasuya and A.~Mazumdar,
  Phys.\ Rev.\  D {\bf 66}, 043505 (2002)
  [arXiv:hep-ph/0206272].


  

\bibitem{Kusenko:2008zm}
  A.~Kusenko and A.~Mazumdar,
  Phys.\ Rev.\ Lett.\  {\bf 101}, 211301 (2008)
  [arXiv:0807.4554 [astro-ph]].
  
\bibitem{Kusenko:2009cv}
  A.~Kusenko, A.~Mazumdar and T.~Multamaki,
  Phys.\ Rev.\  D {\bf 79}, 124034 (2009)
  [arXiv:0902.2197 [astro-ph.CO]].
  
\bibitem{Chiba:2009zu}
  T.~Chiba, K.~Kamada and M.~Yamaguchi,
  Phys.\ Rev.\  D {\bf 81}, 083503 (2010)
  [arXiv:0912.3585 [astro-ph.CO]].
  
\bibitem{Kasuya:2002zs}
  S.~Kasuya, M.~Kawasaki and F.~Takahashi,
  Phys.\ Lett.\  B {\bf 559}, 99 (2003)
  [arXiv:hep-ph/0209358].


\bibitem{Kolb:1993hw}
  E.~W.~Kolb and I.~I.~Tkachev,
  Phys.\ Rev.\  D {\bf 49}, 5040 (1994)
  [arXiv:astro-ph/9311037].

\bibitem{Hiramatsu:2010dx}
  T.~Hiramatsu, M.~Kawasaki and F.~Takahashi,
  JCAP {\bf 1006}, 008 (2010)
  [arXiv:1003.1779 [hep-ph]].


  \bibitem{Gleiser:2010qt}
  M.~Gleiser, N.~Graham and N.~Stamatopoulos,
  arXiv:1004.4658 [astro-ph.CO].

  
\bibitem{Frolov:2008hy}
  A.~V.~Frolov,
  JCAP {\bf 0811}, 009 (2008)
  [arXiv:0809.4904 [hep-ph]].
 
  
\bibitem{Easther:2010qz}
  R.~Easther, H.~Finkel and N.~Roth,
  arXiv:1005.1921 [astro-ph.CO].

\bibitem{Jedamzik:2010dq}
  K.~Jedamzik, M.~Lemoine and J.~Martin,
  arXiv:1002.3039 [astro-ph.CO].
  

\bibitem{Easther:2010mr}
  R.~Easther, R.~Flauger and J.~B.~Gilmore,
  arXiv:1003.3011 [astro-ph.CO].

\bibitem{QFTBook:2007}
  V.~F.~Mukhanov and S.~Winitzki,  {  \em{Introduction to Quantum Effects in Gravity}}
 (Cambridge University Press, Cambridge (2007))
    
\bibitem{Johnson:2008se}
  M.~C.~Johnson and M.~Kamionkowski,
  Phys.\ Rev.\  D {\bf 78}, 063010 (2008)
  [arXiv:0805.1748 [astro-ph]].
  
\bibitem{Guth:1980zm}
  A.~H.~Guth,
  Phys.\ Rev.\  D {\bf 23}, 347 (1981).
  
\bibitem{Linde:1981mu}
  A.~D.~Linde,
  Phys.\ Lett.\  B {\bf 108}, 389 (1982).
  
\bibitem{Albrecht:1982wi}
  A.~Albrecht and P.~J.~Steinhardt,
  Phys.\ Rev.\ Lett.\  {\bf 48}, 1220 (1982).
  
\bibitem{Enqvist:2003gh}
  K.~Enqvist and A.~Mazumdar,
  Phys.\ Rept.\  {\bf 380}, 99 (2003)
  [arXiv:hep-ph/0209244].


\bibitem{Komatsu:2008hk}
  E.~Komatsu {\it et al.}  [WMAP Collaboration],
  Astrophys.\ J.\ Suppl.\  {\bf 180}, 330 (2009)
  [arXiv:0803.0547 [astro-ph]].


\bibitem{Silverstein:2008sg}
  E.~Silverstein and A.~Westphal,
  Phys.\ Rev.\  D {\bf 78}, 106003 (2008)
  [arXiv:0803.3085 [hep-th]].
  
\bibitem{Linde:1993cn}
  A.~D.~Linde,
  Phys.\ Rev.\  D {\bf 49}, 748 (1994)
  [arXiv:astro-ph/9307002].
 

  
\end{thebibliography}
\end{document}